\documentclass[12pt]{article}
\usepackage{amssymb,amsmath,epsfig}
\allowdisplaybreaks

\begin{document}

\title{\bf Effects of $f(\mathcal{R},\mathcal{T},\mathcal{R}_{\gamma\upsilon}\mathcal{T}^{\gamma\upsilon})$
Gravity on Anisotropic Charged Compact Structures}
\author{M. Sharif \thanks{msharif.math@pu.edu.pk}~ and T. Naseer \thanks{tayyabnaseer48@yahoo.com,
tayyab.naseer@math.uol.edu.pk}\\
Department of Mathematics, University of the Punjab,\\
Quaid-i-Azam Campus, Lahore-54590, Pakistan.}

\date{}
\maketitle

\begin{abstract}
This paper focuses on the analysis of static spherically symmetric
anisotropic solutions in the presence of electromagnetic field
through the gravitational decoupling approach in
$f(\mathcal{R},\mathcal{T},\mathcal{R}_{\gamma\upsilon}\mathcal{T}^{\gamma\upsilon})$
gravity. We use geometric deformation only on radial metric function
and obtain two sets of the field equations. The first set deals with
isotropic fluid while the second set yields the influence of
anisotropic source. We consider the modified Krori-Barua charged
isotropic solution for spherical self-gravitating star to deal with
the isotropic system. The second set of the field equations is
solved by taking two different constraints. We then investigate
physical acceptability of the obtained solutions through graphical
analysis of the effective physical variables and energy conditions.
We also analyze the effects of charge on different parameters,
(i.e., mass, compactness and redshift) for the resulting solutions.
It is found that our both solutions are viable as well as stable for
specific values of the decoupling parameter $\varphi$ and charge. We
conclude that a self-gravitating star shows more stable behavior in
this gravity.
\end{abstract}

\section{Introduction}

General Relativity (GR) is considered as one of the main pillars to
understand the fundamental concepts of gravitational phenomena as
well as cosmology. Several cosmological observations suggest that
celestial objects are not dispersed randomly throughout the universe
but are systematically organized. The study of such arrangement of
interstellar objects and their physical properties encourage us to
figure out cosmic expansion. Such expansion is supposed to be
carried out by a mysterious form of energy known as dark energy
which has repulsive nature. Thus the modifications to GR have been
found to be important in revealing hidden facets of our cosmos. The
straightforward modification of GR is $f(\mathcal{R})$ theory, which
is developed by replacing the Ricci scalar $\mathcal{R}$ with its
generic function in the Einstein-Hilbert action. Multiple approaches
have been used by researchers to discuss the feasibility and
stability of this theory \cite{2,3,4,5}. The stability of compact
stars has also been analyzed by Capozziello and his collaborators
\cite{8} through the Lan\'{e}-Emden equation in $f(\mathcal{R})$
theory. Many authors \cite{9,9a,9b,9c,9d,9e,9f,9g} have discussed
the composition and stability of celestial objects in this theory.

To investigate the role of coupling in $f(\mathcal{R})$ gravity,
Bertolami \emph{et al.} \cite{10} assumed the Lagrangian in terms of
$\mathcal{R}$ and $\mathcal{L}_{m}$. This interaction between matter
and geometry in modified theories has prompted many scientists to
concentrate on accelerating interstellar expansion. Harko \emph{et
al.} \cite{20} introduced $f(\mathcal{R},\mathcal{T})$ theory and
studied the effects of non-minimal matter geometry interaction on
self-gravitating structures, $\mathcal{T}$ describes trace of the
energy-momentum tensor. Such an interaction provides
non-conservation of the energy-momentum tensor which may lead to
accelerating expansion of cosmos. Haghani \emph{et al.} \cite{22}
presented a more general interaction so called
$f(\mathcal{R},\mathcal{T},\mathcal{Q})$ gravity, where
$\mathcal{Q}\equiv
\mathcal{R}_{\gamma\upsilon}\mathcal{T}^{\gamma\upsilon}$ to
investigate the effects of strong non-minimal coupling. They
examined the Dolgov-Kawasaki instability and obtained the stability
conditions for this theory. Sharif and Zubair studied viability of
thermodynamical laws \cite{22a} in this theory and evaluated their
energy bounds \cite{22b}.

Odintsov and S\'{a}ez-G\'{o}mez \cite{23} found numerical solutions
of highly complex gravitational equations for some particular models
in this theory and also discussed some issues associated with the
instability of fluid configuration. Ayuso \emph{et al.} \cite{24}
examined the stability of this theory with suitable scalar/vector
fields and concluded that the occurrence of instability is
inevitable in the case of vector field. Baffou \emph{et al.}
\cite{25} obtained field equations for FRW model and analyzed the
viability by constituting perturbation functions. Sharif and Waseem
\cite{25a,25b} discussed physical properties of compact celestial
structures with isotropic and anisotropic matter configurations. In
this framework, Yousaf \emph{et al.} \cite{26,26a,26b,26c,26d,26e}
found four structure scalars which are associated with basic
properties of matter configuration for charged and uncharged
geometries.

The presence of electromagnetic field in self-gravitating structures
plays a significant role in explaining their expansion and
stability. Numerous research has been done in GR as well as modified
theories to investigate the effects of charge on celestial objects.
By matching the static interior isotropic geometry with exterior
Riessner-Nordstr\"{o}m, Das \emph{et al.} \cite{27} found solution
of the Einstein-Maxwell field equations. Sunzu \emph{et al.}
\cite{27a} studied strange stars and observed the effects of charge
through mass-radius relation. Murad \cite{28} developed a charged
model for anisotropic strange star and analyzed its physical
properties. The presence of electric field has been found to
contribute to more stable stellar structures
\cite{28a,28b,28c,28d,28e,28f,28g,28h}.

It has always been an interesting issue to find exact solutions of a
self-gravitating geometry due to non-linear field equations. A
recently developed technique so called minimal geometric deformation
(MGD) through gravitational decoupling has presented its
considerable consequences on achieving physically feasible
solutions. This technique provides several interesting ingredients
for new exact cosmological as well as astrophysical solutions.
Ovalle was the pioneer \cite{29} who suggested this procedure in the
context of braneworld for the development of analytical solutions
corresponding to compact spherical geometries. Ovalle and Linares
\cite{30} constructed exact isotropic solution for spherical
geometry and found these results to be consistent with the Tolman-IV
solution in the braneworld. Casadio \emph{et al.} \cite{31}
developed a new exterior spherical solution which exhibits singular
behavior at Schwarzschild radius.

Ovalle \cite{32} used gravitational decoupling approach and found
exact solution for spherical anisotropic fluid distribution. Ovalle
\emph{et al.} \cite{33} generalized the isotropic solution to new
anisotropic solutions and analyzed their graphical behavior.
Gabbanelli \emph{et al.} \cite{34} evaluated physically acceptable
anisotropic solutions by taking Duragpal-Fuloria isotropic solution.
Estrada and Tello-Ortiz \cite{34a} applied the same method to the
Heintzmann solution and evaluated two physically acceptable
anisotropic solutions. Sharif and Sadiq \cite{35} considered the
Krori-Barua solution and formed charged anisotropic spherical
solutions by using two different constrains. They also investigated
their viability and stability for specific values of the coupling
parameter. This work has been generalized by Sharif and his
collaborators \cite{36,36a,36b,36c} to modified theories such as
$f(\mathcal{G})$ and $f(\mathcal{R})$. Singh \emph{et al.} \cite{37}
utilized class-one condition to construct anisotropic solution via
MGD approach. Hensh and Stuchlik \cite{37a} found physically viable
anisotropic solutions by applying a suitable deformation on Tolman
VII solution. In the Brans-Dicke scenario, Sharif and Majid
\cite{37c,38,39,40,41} found several cosmological solutions by
utilizing minimal as well as extended gravitational decoupling
approach.

This paper examines the effects of modified
$f(\mathcal{R},\mathcal{T},\mathcal{Q})$ corrections on charged
spherical anisotropic solutions developed through MGD scheme. The
article is organized as follows. In the next section, we present the
fundamental formulation of this theory. The MGD technique is
discussed in section 3, which separates the modified field equations
into two sets. We assume the Krori-Barua charged solution in section
4 to find two anisotropic exact solutions. We also discuss
feasibility and stability of the obtained solutions. Finally, our
findings are summarized in section 5.

\section{The $f(\mathcal{R},\mathcal{T},\mathcal{Q})$ Gravity}

In the presence of an additional matter field, the modification of
Einstein-Hilbert action (with $\kappa=8\pi$) is given as \cite{23}
\begin{equation}\label{g1}
S=\int
\left[\frac{f(\mathcal{R},\mathcal{T},\mathcal{R}_{\gamma\upsilon}\mathcal{T}^{\gamma\upsilon})}{16\pi}
+\mathcal{L}_{m}+\mathcal{L}_{\mathcal{E}}+\varphi\mathcal{L}_{\Lambda}\right]\sqrt{-g}d^{4}x,
\end{equation}
where $\mathcal{L}_{m}$ indicates the matter Lagrangian which is
taken to be $-\mu$, $\mu$ is the energy density of fluid
distribution. The Lagrangian densities of electromagnetic field and
new gravitational source are indicated by
$\mathcal{L}_{\mathcal{E}}$ and $\mathcal{L}_{\Lambda}$,
respectively. The corresponding field equations take the form as
\begin{equation}\label{g2}
\mathcal{G}_{\gamma\upsilon}=8\pi
\mathcal{T}_{\gamma\upsilon}^{(tot)}.
\end{equation}
The term $\mathcal{G}_{\gamma\upsilon}$ represents an Einstein
tensor and $\mathcal{T}_{\gamma\upsilon}^{(tot)}$ is the
energy-momentum tensor which can further be written as
\begin{equation}\label{g3}
\mathcal{T}_{\gamma\upsilon}^{(tot)}=\mathcal{T}_{\gamma\upsilon}^{(eff)}+\varphi
\Lambda_{\gamma\upsilon}=\frac{1}{f_{\mathcal{R}}-\mathcal{L}_{m}f_{\mathcal{Q}}}\left(\mathcal{T}_{\gamma\upsilon}^{(m)}
+\mathcal{E}_{\gamma\upsilon}\right)+\mathcal{T}_{\gamma\upsilon}^{(D)}+\varphi\Lambda_{\gamma\upsilon}.
\end{equation}
The new source $\Lambda_{\gamma\upsilon}$ is responsible for
generating effects of anisotropy through scalar, vector or tensor
fields and $\varphi$ is the decoupling parameter which governs its
influence in self-gravitating system. Furthermore,
$\mathcal{T}_{\gamma\upsilon}^{(eff)}$ can be regarded as the
energy-momentum tensor in $f(\mathcal{R},\mathcal{T},\mathcal{Q})$
gravity which encompasses the influence of both usual and modified
corrections. The value of $\mathcal{T}_{\gamma\upsilon}^{(D)}$ in
this case becomes
\begin{eqnarray}
\nonumber
\mathcal{T}_{\gamma\upsilon}^{(D)}&=&-\frac{1}{\mathcal{L}_{m}f_{\mathcal{Q}}-f_{\mathcal{R}}}
\left[\left(f_{\mathcal{T}}+\frac{1}{2}\mathcal{R}f_{\mathcal{Q}}\right)\mathcal{T}_{\gamma\upsilon}^{(m)}
+\left\{\frac{\mathcal{R}}{2}(\frac{f}{\mathcal{R}}-f_{\mathcal{R}})-\mathcal{L}_{m}f_{\mathcal{T}}\right.\right.\\\nonumber
&-&\left.\frac{1}{2}\nabla_{\rho}\nabla_{\eta}(f_{\mathcal{Q}}\mathcal{T}^{\rho\eta})\right\}g_{\gamma\upsilon}
-\frac{1}{2}\Box(f_{\mathcal{Q}}\mathcal{T}_{\gamma\upsilon})-(g_{\gamma\upsilon}\Box-
\nabla_{\gamma}\nabla_{\upsilon})f_{\mathcal{R}}\\\label{g4}
&-&2f_{\mathcal{Q}}\mathcal{R}_{\rho(\gamma}\mathcal{T}_{\upsilon)}^{\rho}+\nabla_{\rho}\nabla_{(\gamma}[\mathcal{T}_{\upsilon)}^{\rho}f_{\mathcal{Q}}]
+2(f_{\mathcal{Q}}\mathcal{R}^{\rho\eta}+\left.f_{\mathcal{T}}g^{\rho\eta})\frac{\partial^2\mathcal{L}_{m}}
{\partial g^{\gamma\upsilon}\partial g^{\rho\eta}}\right],
\end{eqnarray}
where $f_{\mathcal{R}}=\frac{\partial
f(\mathcal{R},\mathcal{T},\mathcal{Q})}{\partial
\mathcal{R}},~f_{\mathcal{T}}=\frac{\partial
f(\mathcal{R},\mathcal{T},\mathcal{Q})}{\partial
\mathcal{T}},~f_{\mathcal{Q}}=\frac{\partial
f(\mathcal{R},\mathcal{T},\mathcal{Q})}{\partial
\mathcal{Q}},~\nabla_\rho$ symbolizes the covariant derivative and
$\Box\equiv g^{\gamma\upsilon}\nabla_\gamma\nabla_\upsilon$.

For a perfect matter configuration, the energy-momentum tensor is
\begin{equation}\label{g5}
\mathcal{T}_{\gamma\upsilon}^{(m)}=(\mu+P) \mathcal{U}_{\gamma}
\mathcal{U}_{\upsilon}+Pg_{\gamma\upsilon},
\end{equation}
where $\mathcal{U}_{\gamma}$ and $P$ represent the four-velocity and
isotropic pressure, repectively. A distinctive relationship between
$\mathcal{R}$ and $\mathcal{T}$ has been presented by the trace of
energy-momentum tensor in GR. By taking the trace of modified field
equations of $f(\mathcal{R},\mathcal{T},\mathcal{Q})$ theory, one
can have the following form as
\begin{align}\nonumber
&3\nabla^{\upsilon}\nabla_{\upsilon}
f_\mathcal{R}+\mathcal{R}\left(f_\mathcal{R}-\frac{\mathcal{T}}{2}f_\mathcal{Q}\right)-\mathcal{T}(f_\mathcal{T}+1)+\frac{1}{2}
\nabla^{\upsilon}\nabla_{\upsilon}(f_\mathcal{Q}\mathcal{T})+\nabla_\gamma\nabla_\upsilon(f_\mathcal{Q}\mathcal{T}^{\gamma\upsilon})\\\nonumber
&-2f+(\mathcal{R}f_\mathcal{Q}+4f_\mathcal{T})\mathcal{L}_m+2\mathcal{R}_{\gamma\upsilon}\mathcal{T}^{\gamma\upsilon}f_\mathcal{Q}
-2g^{\rho\eta} \frac{\partial^2\mathcal{L}_m}{\partial
g^{\rho\eta}\partial
g^{\gamma\upsilon}}\left(f_\mathcal{T}g^{\gamma\upsilon}+f_\mathcal{Q}R^{\gamma\upsilon}\right)=0.
\end{align}
By considering $\mathcal{Q}=0$ in the above equation, one can obtain
$f(\mathcal{R},\mathcal{T})$ theory, while $f(\mathcal{R})$ theory
can be retrieved with the consideration of vacuum case. The
energy-momentum tensor for electromagnetic field has the form
\begin{equation*}
\mathcal{E}_{\gamma\upsilon}=\frac{1}{4\pi}\left[\frac{1}{4}g_{\gamma\upsilon}\mathcal{F}^{\rho\eta}\mathcal{F}_{\rho\eta}
-\mathcal{F}^{\eta}_{\gamma}\mathcal{F}_{\upsilon\eta}\right],
\end{equation*}
where
$\mathcal{F}_{\gamma\upsilon}=\alpha_{\upsilon;\gamma}-\alpha_{\gamma;\upsilon}$
is the Maxwell field tensor and
$\alpha_{\upsilon}=\alpha(r)\delta^{\upsilon}_{0}$ indicates the
four potential, which satisfies the following equations of tensorial
form
\begin{equation*}
\mathcal{F}^{\gamma\upsilon}_{;\upsilon}=4\pi \jmath^{\gamma}, \quad
F_{[\gamma\upsilon;\eta]}=0,
\end{equation*}
$\jmath^{\gamma}$ is the current density which can be written in
terms of charge density $\varrho$ as $\jmath^{\gamma}=\varrho
\mathcal{U}^{\gamma}$.

We consider a hypersurface $\Sigma$ which separates the interior and
exterior regions of our geometry. The spherical cosmic object which
is static in nature can be defined by the metric as
\begin{equation}\label{g6}
ds^2=-e^{\xi} dt^2+e^{\chi} dr^2+r^2d\theta^2+r^2\sin^2\theta
d\phi^2,
\end{equation}
where $\xi=\xi(r)$ and $\chi=\chi(r)$. The Maxwell field equations
for the above metric takes the form
\begin{equation}
\alpha''+\frac{1}{2r}\left(4-r(\xi'+\chi')\right)\alpha'=4\pi\varrho
e^{\frac{\xi}{2}+\chi},
\end{equation}
whose integration provides
\begin{equation}
\alpha'=\frac{q}{r^2}e^{\frac{\xi+\chi}{2}},
\end{equation}
where $q$ indicates charge in the inner region of self-gravitating
body. Here, $'=\frac{\partial}{\partial r}$.

We consider a comoving frame, so the four-velocity and four-vector
are
\begin{equation}\label{g7}
\mathcal{U}^\upsilon=(e^{\frac{-\xi}{2}},0,0,0),\quad
\mathcal{W}^\upsilon=(0,e^{\frac{-\chi}{2}},0,0),
\end{equation}
with the relations $\mathcal{U}^\upsilon \mathcal{U}_{\upsilon}=-1$
and $\mathcal{W}^\upsilon \mathcal{U}_{\upsilon}=0$. For spherically
symmetric spacetime, the field equations in
$f(\mathcal{R},\mathcal{T},\mathcal{Q})$ theory are
\begin{align}\label{g8}
&e^{-\chi}\left(\frac{\chi'}{r}-\frac{1}{r^2}\right)
+\frac{1}{r^2}=8\pi\left(\tilde{\mu}+\frac{s^2}{8\pi
r^4}-\mathcal{T}_{0}^{0(\mathcal{D})}-\varphi
\Lambda_{0}^{0}\right),\\\label{g9}
&e^{-\chi}\left(\frac{\xi'}{r}+\frac{1}{r^2}\right)
-\frac{1}{r^2}=8\pi\left(\tilde{P}-\frac{s^2}{8\pi
r^4}+\mathcal{T}_{1}^{1(\mathcal{D})}+\varphi\Lambda_{1}^{1}\right),
\\\label{g10}
&\frac{e^{-\chi}}{4}\left[2\xi''+\xi'^2-\xi'\chi'+\frac{2\xi'}{r}-\frac{2\chi'}{r}\right]
=8\pi\left(\tilde{P}+\frac{s^2}{8\pi
r^4}+\mathcal{T}_{2}^{2(\mathcal{D})}+\varphi\Lambda_{2}^{2}\right),
\end{align}
where $\tilde{\mu}=\frac{1}{(f_{\mathcal{R}}+\mu
f_{\mathcal{Q}})}\mu$,~$\tilde{P}=\frac{1}{(f_{\mathcal{R}}+\mu
f_{\mathcal{Q}})}P$ and $s^2=\frac{1}{(f_{\mathcal{R}}+\mu
f_{\mathcal{Q}})}q^2$. The above field equations become highly
complicated due to the presence of correction terms
$\mathcal{T}_{0}^{0(\mathcal{D})},~\mathcal{T}_{1}^{1(\mathcal{D})}$
and $\mathcal{T}_{2}^{2(\mathcal{D})}$ of modified gravity whose
values are presented in Appendix \textbf{A}.

In contrast to GR and $f(\mathcal{R})$ theory, the stress-energy
tensor in this theory has non-zero divergence, (i.e., $\nabla^\gamma
\mathcal{T}_{\gamma\upsilon}\neq 0$) due to the interaction between
matter and geometry which leads to a violation of the equivalence
principle. Thus the moving molecules do not obey geodesic path in
the gravitational field due to the existence of an additional force.
Hence we have
\begin{align}\nonumber
\nabla^\gamma
\mathcal{T}_{\gamma\upsilon}&=\frac{2}{2f_\mathcal{T}+\mathcal{R}f_\mathcal{Q}+1}\left[\nabla_\upsilon(\mathcal{L}_mf_\mathcal{T})
+\nabla_\gamma(f_\mathcal{Q}\mathcal{R}^{\rho\gamma}\mathcal{T}_{\rho\upsilon})-\frac{1}{2}
(f_\mathcal{T}g_{\rho\eta}+f_\mathcal{Q}\mathcal{R}_{\rho\eta})\right.\\\label{g11}
&\times\left.\nabla_\upsilon
\mathcal{T}^{\rho\eta}-\mathcal{G}_{\gamma\upsilon}\nabla^\gamma(f_\mathcal{Q}\textit{L}_m)\right].
\end{align}
This induces the hydrostatic equilibrium condition as
\begin{align}\nonumber
&\frac{d\tilde{P}}{dr}+\frac{d\mathcal{T}_{1}^{1(\mathcal{D})}}{dr}+\varphi
\frac{d\Lambda_{1}^{1}}{dr}-\frac{ss'}{4\pi
r^4}+\frac{\xi'}{2}\left(\tilde{\mu}
+\tilde{P}+\mathcal{T}_{1}^{1(\mathcal{D})}-\mathcal{T}_{0}^{0(\mathcal{D})}\right)-\frac{2\varphi}{r}
\left(\Lambda_{2}^{2}-\Lambda_{1}^{1} \right)\\\label{g12}
&+\frac{2}{r}\left(\mathcal{T}_{1}^{1(\mathcal{D})}-\mathcal{T}_{2}^{2(\mathcal{D})}\right)-\frac{\varphi\xi'}{2}
\left(\Lambda_{0}^{0}-\Lambda_{1}^{1}\right)=\Omega,
\end{align}
where $\Omega$ presents the dark source terms of modified gravity
and its value is casted in Appendix \textbf{A}. Equation \eqref{g12}
can be seen as the generalization of Tolman-Opphenheimer-Volkoff
(TOV) equation. This equation seems to be very important in order to
explain the systematic revolutions in self-gravitating spherical
structures.

We have a set of highly non-linear differential equations
\eqref{g8}-\eqref{g10} and \eqref{g12}, which contains eight
unknowns
$(\xi,\chi,\mu,P,s,\Lambda_{0}^{0},\Lambda_{1}^{1},\Lambda_{2}^{2})$,
hence, this system is no longer definite. To close the above system,
we use systematic approach \cite{33} and then determine the unknown
quantities. The effective matter variables for the field equations
\eqref{g8}-\eqref{g10} can be found as
\begin{equation}\label{g13}
\hat{\mu}=\tilde{\mu}-\varphi\Lambda_{0}^{0},\quad
\hat{P}_{r}=\tilde{P}+\varphi\Lambda_{1}^{1}, \quad
\hat{P}_{\bot}=\tilde{P}+\varphi\Lambda_{2}^{2}.
\end{equation}
This shows that the source $\Lambda_{\gamma}^{\upsilon}$ induces
anisotropy within a self-gravitating structure. Thus we define the
effective anisotropy as
\begin{equation}\label{g14}
\hat{\Delta}=\hat{P}_{\bot}-\hat{P}_{r}=\varphi\left(\Lambda_{2}^{2}-\Lambda_{1}^{1}\right),
\end{equation}
which does not appear for $\varphi=0$.

\section{Gravitational Decoupling}

In this section, we adopt a new procedure known as gravitational
decoupling via MGD technique to solve the system
\eqref{g8}-\eqref{g10} and determine unknowns. This approach
transforms the field equations in such a manner that the additional
source $\Lambda_{\gamma}^{\upsilon}$ provides effective equations
which may generate anisotropy in the interior of the star. We start
with basic constituent of this technique which is perfect fluid
solution $(\omega,\nu,\mu,P,s)$ with the metric
\begin{equation}\label{g15}
ds^2=-e^{\omega}dt^2+\frac{1}{\nu}dr^2+r^2d\theta^2+r^2\sin^2\theta
d\phi^2,
\end{equation}
where $\omega=\omega(r)$ and
$\nu=\nu(r)=1-\frac{2m(r)}{r}+\frac{s^2}{r^2}$, $m(r)$ denotes the
Misner-Sharp mass of the star having perfect matter distribution.
Next, to incorporate the effects of new source
$\Lambda_{\gamma}^{\upsilon}$ on charged isotropic solution, we
impose the linear form of geometrical transformations on the metric
components as
\begin{equation}\label{g16}
\omega\rightarrow\xi=\omega+\varphi a, \quad \nu\rightarrow
e^{-\chi}=\nu+\varphi b,
\end{equation}
where the two translation functions $a$ and $b$ correspond to
temporal and radial metric coefficients, respectively. The MGD
approach only ensures the translation of radial component under the
influence of additional source, while the temporal component remains
unchanged. Thus we have $a=0,~b\rightarrow b^*$. Equation
\eqref{g16} is then reduced to
\begin{equation}\label{g17}
\omega\rightarrow\xi=\omega, \quad \nu\rightarrow
e^{-\chi}=\nu+\varphi b^*,
\end{equation}
where $b^*=b^*(r)$.

We split the field equations into two sets to solve the complex
system. After applying the above transformations in
Eqs.\eqref{g8}-\eqref{g10}, we obtain the first set of differential
equations for $\varphi=0$ as
\begin{align}\label{g18}
&8\pi\left(\tilde{\mu}+\frac{s^2}{8\pi
r^4}-\mathcal{T}_{0}^{0(\mathcal{D})}\right)=\frac{1}{r^2}-\left(\frac{\nu'}{r}+\frac{\nu}{r^2}\right),\\\label{g19}
&8\pi\left(\tilde{P}-\frac{s^2}{8\pi
r^4}+\mathcal{T}_{1}^{1(\mathcal{D})}\right)=-\frac{1}{r^2}+\nu\left(\frac{\xi'}{r}+\frac{1}{r^2}\right),\\\label{g20}
&8\pi\left(\tilde{P}+\frac{s^2}{8\pi
r^4}+\mathcal{T}_{2}^{2(\mathcal{D})}\right)=\nu\left(\frac{\xi''}{2}+\frac{\xi'^2}{4}+\frac{\xi'}{2r}\right)+\nu'\left(\frac{\xi'}{4}
+\frac{1}{2r}\right).
\end{align}
The second set encompasses the source $\Lambda^{\upsilon}_{\gamma}$
has the form
\begin{align}\label{g21}
&8\pi\Lambda_{0}^{0}=\frac{{b^*}'}{r}+\frac{b^*}{r^2},\\\label{g22}
&8\pi\Lambda_{1}^{1}=b^*\left(\frac{\xi'}{r}+\frac{1}{r^2}\right),\\\label{g23}
&8\pi\Lambda_{2}^{2}=b^*\left(\frac{\xi''}{2}+\frac{\xi'^2}{4}+\frac{\xi'}{2r}\right)+{b^*}'\left(\frac{\xi'}{4}
+\frac{1}{2r}\right).
\end{align}
The set of equations \eqref{g21}-\eqref{g23} seems to be equivalent
to the field equations for anisotropic charged spherical stellar
configuration with matter variables defined as
$\hat{\mu}=\Lambda_{0}^{0},~\hat{P}_{r}=-\Lambda_{1}^{1},~\hat{P}_{\bot}=-\Lambda_{2}^{2}$
signifying the metric
\begin{equation}\label{g24}
ds^2=-e^{\xi}dt^2+\frac{1}{b^*}dr^2+r^2d\theta^2+r^2\sin^2\theta
d\phi^2.
\end{equation}
Equations \eqref{g21}-\eqref{g23} differ from isotropic field
equations for charged sphere by a single term $\frac{1}{r^2}$. Thus,
to make this set as the standard field equations for anisotropic
fluid, we define matter components as $\hat{\mu}+\frac{s^2}{8\pi
r^4}=\Lambda_{0}^{*0}=\Lambda_{0}^{0}+\frac{1}{r^2},\quad
\hat{P}_{r}-\frac{s^2}{8\pi
r^4}=\Lambda_{1}^{*1}=\Lambda_{1}^{1}+\frac{1}{r^2},\quad
\hat{P}_{\bot}+\frac{s^2}{8\pi
r^4}=\Lambda_{2}^{*2}=\Lambda_{2}^{2}=\Lambda_{3}^{*3}=\Lambda_{3}^{3}$.
Hence, the MGD approach has transformed the indefinite system
\eqref{g8}-\eqref{g10} into two sets of equations. The first set
\eqref{g18}-\eqref{g20} describes the isotropic fluid
$(\tilde{\mu},\tilde{P},\xi,\chi$) whereas the second set
\eqref{g21}-\eqref{g23} obeys the above anisotropic system and
contains four unknowns
($b^*,\Lambda_{0}^{0},\Lambda_{1}^{1},\Lambda_{2}^{2}$). As a
consequence, we successfully decoupled the system
\eqref{g8}-\eqref{g10}.

In order to investigate basic features of the celestial bodies, some
constraints on the boundary surface $(\Sigma)$ play a key role,
known as junction conditions. These conditions allow a smooth
matching of inner and outer regions of the star at $\Sigma$. For
this purpose, the inner geometry is given as
\begin{equation}\label{g25}
ds^2=-e^{\xi}dt^2+\frac{1}{\left(1-\frac{2\check{m}(r)}{r}+\frac{s^2}{r^2}\right)}dr^2+
r^2d\theta^2+r^2\sin^2\theta d\phi^2,
\end{equation}
where $\check{m}(r)=m(r)-\frac{\varphi r}{2}b^*(r)$ is the interior
mass. For matching the two geometries (inner and outer), we need an
outer spacetime of the star given by
\begin{equation}\label{g26}
ds^2=-e^{\xi} dt^2+e^{\chi} dr^2+r^2d\theta^2+r^2\sin^2\theta
d\phi^2.
\end{equation}
The first fundamental form $([ds^2]_{\Sigma}=0)$ of the junction
conditions at hypersurface provides
\begin{equation}\label{g27}
\xi_{-}(R)=\xi_{+}(R), \quad
1-e^{-\chi_{+}(R)}=\frac{2M_{0}}{R}-\frac{S^2_{0}}{R^2}- \varphi
b^*(R),
\end{equation}
where $\nu=e^{-\chi}-\varphi b^*$. Here, $-$ and $+$ signs in
subscript correspond to the metric coefficients of inner and outer
geometry, respectively. Also, $M_{0}=m(R),~S^2_{0}=s^2(R)$ and
$b^*(R)$ indicate the total mass, charge and deformation at the
boundary. Moreover, the second form
($[\mathcal{T}^{(tot)}_{\gamma\upsilon}\mathcal{W}^{\upsilon}]_{\Sigma}=0$,
$\mathcal{W}^{\upsilon}$ is defined in Eq.\eqref{g7}) yields
\begin{equation}\label{g28}
\tilde{P}(R)-\frac{S_{0}^{2}}{8\pi R^4}+\varphi
\left(\Lambda^{1}_{1}(R)\right)_{-}+
\left(\mathcal{T}^{1(\mathcal{D})}_{1}(R)\right)_{-}=\varphi
\left(\Lambda^{1}_{1}(R)\right)_{+}+\left(\mathcal{T}^{1(\mathcal{D})}_{1}(R)\right)_{+}.
\end{equation}
By making use of Eq.\eqref{g27}, the above equation takes the form
\begin{equation}\label{g29}
\tilde{P}(R)-\frac{S_{0}^{2}}{8\pi R^4}+\varphi
\left(\Lambda^{1}_{1}(R)\right)_{-}=\varphi
\left(\Lambda^{1}_{1}(R)\right)_{+},
\end{equation}
which can also be written as
\begin{equation}\label{g30}
\tilde{P}(R)-\frac{S_{0}^{2}}{8\pi R^4}+\frac{\varphi
b^*(R)}{8\pi}\left(\frac{\xi'(R)}{R}+
\frac{1}{R^2}\right)=\frac{\varphi g^*(R)}{8\pi
R^2}\left(\frac{R^2-\mathcal{S}^2}{R^2-2\mathcal{M}R+\mathcal{S}^2}\right),
\end{equation}
where $\mathcal{M}$ and $\mathcal{S}$ show the mass and charge of
the outer geometry, respectively. The term $g^*$ denotes the
exterior radial geometric deformation for the Riessner-Nordstr\"{o}m
metric in the presence of $\Lambda_{\gamma\upsilon}$ (source)
desribed by
\begin{equation}\label{g31}
ds^2=-\left(1-\frac{2\mathcal{M}}{r}+\frac{\mathcal{S}^2}{r^2}\right)dt^2+\frac{1}{\left(1-\frac{2\mathcal{M}}{r}+\frac{\mathcal{S}^2}{r^2}
+\varphi g^*\right)}dr^2+ r^2d\theta^2+r^2\sin^2\theta d\phi^2.
\end{equation}
The MGD inner and outer Riessner-Nordstr\"{o}m geometries (both are
filled with source $\Lambda_{\gamma\upsilon}$) are interlinked
through some appropriate conditions provided by two equations
\eqref{g27} and \eqref{g30}. However, by taking the standard
Riessner-Nordstr\"{o}m metric, (i.e., $g^*=0$) as exterior
spacetime, Eq.\eqref{g30} becomes
\begin{equation}\label{g32}
\hat{P}(R)-\frac{S_{0}^{2}}{8\pi R^4}\equiv
\tilde{P}(R)-\frac{S_{0}^{2}}{8\pi R^4}+\frac{\varphi b^*(R)}
{8\pi}\left(\frac{\xi'}{R}+\frac{1}{R^2}\right)=0.
\end{equation}

\section{Anisotropic Solutions}

Our goal is to achieve anisotropic solutions for charged compact
star via MGD approach. To this end, we need charged isotropic
solution of the field equations in
$f(\mathcal{R},\mathcal{T},\mathcal{Q})$ scenario. We assume the
Krori-Barua solution \cite{42} which is very popular due to its
non-singular nature. In the background of this theory, the solution
becomes
\begin{eqnarray}\label{g33}
e^{\xi}&=&e^{\mathcal{A}r^2+\mathcal{B}}, \\\label{g34}
e^{\chi}&=&\nu^{-1}=e^{\mathcal{C}r^2}, \\\nonumber
\tilde{\mu}&=&\frac{1}{16\pi}\left[e^{-\mathcal{C}r^2}
\left\{\mathcal{A}r^2(\mathcal{C}-\mathcal{A})+5\mathcal{C}-\frac{1}{r^2}\right\}+\frac{1}{r^2}\right]+\mathcal{T}^{0(\mathcal{D})}_{0}\\\label{g35}
&-&\frac{1}{2}\left(\mathcal{T}^{1(\mathcal{D})}_{1}-\mathcal{T}^{2(\mathcal{D})}_{2}\right),\\\nonumber
\tilde{P}&=&\frac{1}{16\pi}\left[e^{-\mathcal{C}r^2}\left\{4\mathcal{A}-\mathcal{A}r^2(\mathcal{C}-\mathcal{A})
-\mathcal{C}+\frac{1}{r^2}\right\}-\frac{1}{r^2}\right]\\\label{g36}
&-&\frac{1}{2}\left(\mathcal{T}^{1(\mathcal{D})}_{1}+\mathcal{T}^{2(\mathcal{D})}_{2}\right),\\\nonumber
s^2&=&\frac{r^2}{2}\left[e^{-\mathcal{C}r^2}\left\{\mathcal{A}r^4(\mathcal{A}-\mathcal{C})
-\mathcal{C}r^2-1\right\}+1\right]\\\label{g36a} &+&4\pi
r^4\left(\mathcal{T}^{1(\mathcal{D})}_{1}-\mathcal{T}^{2(\mathcal{D})}_{2}\right),
\end{eqnarray}
where the constants $\mathcal{A},~\mathcal{B}$ and $\mathcal{C}$ are
unknown and can be determined through junction conditions. The
continuity of metric coefficients $(g_{tt},~g_{rr}$ and $g_{tt,r})$
between the interior and exterior geometries produces these unknowns
as
\begin{eqnarray}\label{g37}
\mathcal{A}&=&\frac{1}{R^2}\left(\frac{M_{0}}{R}-\frac{S_{0}^{2}}{R^2}\right)
\left(1-\frac{2M_{0}}{R}+\frac{S_{0}^{2}}{R^2}\right)^{-1},\\\label{g37a}
\mathcal{B}&=&\ln\left(1-\frac{2M_{0}}{R}+\frac{S_{0}^{2}}{R^2}\right)
-\frac{M_{0}R-S^{2}_{0}}{R^2-2M_{0}R+S^{2}_{0}},\\\label{g38}
\mathcal{C}&=&-\frac{1}{R^2}
\ln\left(1-\frac{2M_{0}}{R}+\frac{S_{0}^{2}}{R^2}\right),
\end{eqnarray}
with compactness factor $\frac{2M_{0}}{R}<\frac{8}{9}$. At the
boundary $(r=R)$ of the star, these expressions guarantee the
continuity of the charged isotropic solution
\eqref{g33}-\eqref{g36a} with the outer Riessner-Nordstr\"{o}m
geometry, which is likely to be modified in the inner region with
the inclusion of additional source $\Lambda_{\gamma\upsilon}$. With
the help of temporal and radial metric functions presented in
Eqs.\eqref{g33} and \eqref{g17}, respectively, the anisotropic
solution ($\varphi\neq 0$) in the inner spherical geometry can be
constructed. Equations \eqref{g21}-\eqref{g23} provide the relation
between source $\Lambda_{\gamma\upsilon}$ and geometric deformation
$b^*$, whose solution can be determined by assuming some
supplementary constraints. In the following, we introduce some
conditions to produce two physically feasible anisotropic solutions.

\subsection{Solution I}

Here, by taking a constraint on $\Lambda_{1}^{1}$, we evaluate both
$b^*$ and $\Lambda_{\gamma}^{\upsilon}$ which will be used to find
solution of the anisotropic field equations. It can be observed from
Eq.\eqref{g32} that the exterior Riessner-Nordstr\"{o}m geometry is
found to be compatible with interior isotropic spacetime as long as
$\tilde{P}(R)-\frac{S_{0}^{2}}{8\pi
R^4}+\mathcal{T}_{1}^{1(\mathcal{D})}(R)\sim \varphi
\left(\Lambda_{1}^{1}(R)\right)_{-}$. To meet this requirement, the
easiest choice is to take \cite{33}
\begin{equation}\label{g39}
\tilde{P}-\frac{s^{2}}{8\pi
r^4}+\mathcal{T}_{1}^{1(\mathcal{D})}=\Lambda_{1}^{1}
\quad\Rightarrow\quad b^*=\nu-\frac{1}{1+\xi'r}.
\end{equation}
where Eqs.\eqref{g19} and \eqref{g22} are used. Using
Eq.\eqref{g39}, we obtain radial metric function as
\begin{equation}\label{g40}
e^{-\chi}=(1+\varphi)\nu-\frac{\varphi}{1+2\mathcal{A}r^2}.
\end{equation}
The minimally deformed Krori-Barua solution through anisotropic
source $\Lambda_{\gamma\upsilon}$ can be illustrated by the metric
coefficients given in Eqs.\eqref{g33} and \eqref{g40}. Equation
\eqref{g40} reduces to the standard solution for a compact sphere by
taking $\varphi\rightarrow 0$. The first fundamental form of the
junction conditions provides
\begin{equation}\label{g41}
R^2e^{\mathcal{A}R^2+\mathcal{B}}=R^2-2\mathcal{M}R+\mathcal{S}^2,
\end{equation}
and
\begin{equation}\label{g42}
(1+\varphi)\nu-\frac{\varphi}{1+2\mathcal{A}R^2}=1-\frac{2\mathcal{M}}{R}+\frac{\mathcal{S}^2}{R^2}.
\end{equation}
In the same way, the second fundamental form
($\tilde{P}(R)-\frac{S_{0}^{2}}{8\pi
R^2}+\mathcal{T}_{1}^{1(\mathcal{D})}(R)+\varphi
\left(\Lambda_{1}^{1}(R)\right)_{-}=0$) together with Eq.\eqref{g39}
gives
\begin{equation}\label{g43}
\tilde{P}(R)-\frac{S_{0}^{2}}{8\pi
R^2}+\mathcal{T}_{1}^{1(\mathcal{D})}(R)=0 \quad\Rightarrow\quad
\mathcal{C}=\frac{\ln\left(1+2\mathcal{A}R^2\right)}{R^2}.
\end{equation}
Also, the expression for mass can be found by using Eq.\eqref{g42}
as
\begin{equation}\label{g44}
2R\left(\mathcal{M}-M_{0}\right)=\frac{\varphi
R^2}{1+2\mathcal{A}R^2}+\mathcal{S}^2-S_{0}^{2}-\varphi\left(R^2-2M_{0}R+S_{0}^{2}\right).
\end{equation}
By inserting the value of $\mathcal{M}$ from the above equation in
Eq.\eqref{g41}, we have
\begin{equation}\label{g45}
\mathcal{A}R^2+\mathcal{B}=\ln\left\{\left(1+\varphi\right)\left(1-\frac{2M_{0}}{R}+\frac{S_{0}^2}{R^2}\right)
-\frac{\varphi}{1+2\mathcal{A}R^2}\right\}.
\end{equation}
This equation relates the two constants $\mathcal{A}$ and
$\mathcal{B}$. The system of equations \eqref{g43}-\eqref{g45}
provide some appropriate conditions which are found to be helpful in
order to match the inner and outer geometries smoothly. Finally, by
employing the constraint \eqref{g39}, the expressions for charged
anisotropic solution ($\hat{\mu},\hat{P}_{r},\hat{P}_{\bot},s$) and
anisotropic factor ($\hat{\Delta}$) take the form
\begin{eqnarray}\nonumber
\hat{\mu}&=&\frac{e^{-\mathcal{C}r^2}}{16\pi
r^2}\left[\mathcal{A}r^4(\mathcal{C}-\mathcal{A})+\mathcal{C}r^2(5+4\varphi)-1-2\varphi\right]
+\frac{1}{16\pi r^2\left(1+2\mathcal{A}r^2\right)^2}\\\label{g46}
&\times&\left[1+4\mathcal{A}r^2(1-\varphi)+4\mathcal{A}^2r^4+2\varphi\right]+\mathcal{T}^{0(\mathcal{D})}_{0}
-\frac{1}{2}\left(\mathcal{T}^{1(\mathcal{D})}_{1}-\mathcal{T}^{2(\mathcal{D})}_{2}\right),\\\nonumber
\hat{P}_{r}&=&\frac{1}{16\pi r^2}\left[e^{-\mathcal{C}r^2}
\left\{4\mathcal{A}r^2(1+\varphi)+1-\mathcal{C}r^2-\mathcal{A}r^4(\mathcal{C}-\mathcal{A})+2\varphi\right\}-1-2\varphi\right]\\\label{g47}
&-&\frac{1}{2}\left(\mathcal{T}^{1(\mathcal{D})}_{1}+\mathcal{T}^{2(\mathcal{D})}_{2}\right),\\\nonumber
\hat{P}_{\bot}&=&\frac{1}{16\pi r^2}\left[e^{-\mathcal{C}r^2}
\{4\mathcal{A}r^2(1+\varphi)+1+r^2(\mathcal{A}^2r^2-\mathcal{A}\mathcal{C}r^2-\mathcal{C})(1+2\varphi)\}-1\right]\\\label{g48}
&-&\frac{\varphi}{8\pi\left(1+2\mathcal{A}r^2\right)}
\left[2\mathcal{A}+\mathcal{A}r^2(\mathcal{A}-\mathcal{C})-\mathcal{C}\right]
-\frac{1}{2}\left(\mathcal{T}^{1(\mathcal{D})}_{1}+\mathcal{T}^{2(\mathcal{D})}_{2}\right),\\\nonumber
s^2&=&\frac{r^2}{2}\left[\mathcal{A}r^4(\mathcal{A}-\mathcal{C})-\mathcal{C}r^2-1\right]\left[1+(1-\varphi)e^{-\mathcal{C}r^2}
+\frac{\varphi}{1+2\mathcal{A}r^2}\right]\\\label{g48a} &+&4\pi
r^4\left(\mathcal{T}^{1(\mathcal{D})}_{1}
-\mathcal{T}^{2(\mathcal{D})}_{2}\right),\\\label{g49}
\hat{\Delta}&=&\frac{\varphi}{8\pi r^2}\left[\mathcal{A}r^4
(\mathcal{A}-\mathcal{C})-\mathcal{C}r^2-1\right]\left(e^{-\mathcal{C}r^2}-\frac{1}{1+2\mathcal{A}r^2}\right).
\end{eqnarray}

\subsection{Solution II}

In this subsection, we assume density like constraint to achieve
another physically viable solution for charged anisotropic
configuration. This is given as
\begin{equation}\label{g51}
\tilde{\mu}+\frac{s^2}{8\pi
r^4}-\mathcal{T}_{0}^{0(\mathcal{D})}=\Lambda_{0}^{0},
\end{equation}
making use of Eqs.\eqref{g18} and \eqref{g21}, this becomes
\begin{equation}\label{g52}
{b^*}'+\frac{b^*}{r}-\frac{1}{r}\left[e^{-\mathcal{C}r^2(2\mathcal{C}r^2-1)}+1\right]=0,
\end{equation}
whose integration leads to
\begin{equation}\label{g53}
b^*=\frac{\mathfrak{e}_{1}}{r}+e^{-\mathcal{C}r^2}-1,
\end{equation}
where $\mathfrak{e}_{1}$ is the integration constant. We suppose
$\mathfrak{e}_{1}=0$ to obtain a non-singular solution at the core
$(r=0)$ of a star so that
\begin{equation}\label{g54}
b^*=e^{-\mathcal{C}r^2}-1.
\end{equation}
Using the same strategy as for the solution-I, we can find the
matching conditions for this solution as follows
\begin{eqnarray}\label{g55}
&&2\left(\mathcal{M}-M_{0}\right)R-\mathcal{S}^2+S_{0}^{2}+\varphi
R^2\left(e^{-\mathcal{C}R^2}-1\right)=0,\\\label{g56}
&&\mathcal{A}R^2+\mathcal{B}=\ln\left[1-\frac{2M_{0}}{R}+\frac{S_{0}^{2}}{R^2}+\varphi
\left(e^{-\mathcal{C}R^2}-1\right)\right].
\end{eqnarray}
Finally, the expressions of charged anisotropic solution for density
like constraint \eqref{g51} are obtained as
\begin{align}\nonumber
\hat{\mu}&=\frac{1}{16\pi
r^2}\left[e^{-\mathcal{C}r^2}\left\{\mathcal{A}r^4(\mathcal{C}-\mathcal{A})+\mathcal{C}r^2(5+4\varphi)
-1-2\varphi\right\}+1+2\varphi\right]+\mathcal{T}^{0(\mathcal{D})}_{0}\\\label{g57}
&-\frac{1}{2}\left(\mathcal{T}^{1(\mathcal{D})}_{1}-\mathcal{T}^{2(\mathcal{D})}_{2}\right),\\\nonumber
\hat{P}_{r}&=\frac{1}{16\pi r^2}\left[e^{-\mathcal{C}r^2}
\left\{4\mathcal{A}r^2(1+\varphi)-\mathcal{A}r^4(\mathcal{C}-\mathcal{A})-\mathcal{C}r^2+1+2\varphi\right\}-1-2\varphi\right.\\\label{g58}
&-\left.4\varphi \mathcal{A}r^2\right]
-\frac{1}{2}\left(\mathcal{T}^{1(\mathcal{D})}_{1}+\mathcal{T}^{2(\mathcal{D})}_{2}\right),\\\nonumber
\hat{P}_{\bot}&=\frac{1}{16\pi r^2}\left[e^{-\mathcal{C}r^2}
\{4\mathcal{A}r^2(1+\varphi)+r^2(\mathcal{A}^2r^2-\mathcal{A}\mathcal{C}r^2-\mathcal{C})(1+2\varphi)+1\}-1\right.\\\label{g59}
&-\left.2\varphi r^2\left\{2\mathcal{A}-\mathcal{A}r^2(\mathcal{C}
-\mathcal{A})-\mathcal{C}\right\}\right]
-\frac{1}{2}\left(\mathcal{T}^{1(\mathcal{D})}_{1}+\mathcal{T}^{2(\mathcal{D})}_{2}\right),\\\nonumber
s^2&=\frac{r^2(1-\varphi)}{2}\left[1+e^{-\mathcal{C}r^2}\left\{\mathcal{A}r^4(\mathcal{A}-\mathcal{C})
-\mathcal{C}r^2-1\right\}\right]+\frac{\varphi r^4}{2}\\\label{g59a}
&\times\left(\mathcal{A}^2r^2
-\mathcal{A}\mathcal{C}r^2-\mathcal{C}\right)+4\pi r^4
\left(\mathcal{T}^{1(\mathcal{D})}_{1}-\mathcal{T}^{2(\mathcal{D})}_{2}\right),\\\label{g60}
\hat{\Delta}&=\frac{\varphi}{8\pi r^2}\left[\mathcal{A}r^4
(\mathcal{A}-\mathcal{C})-\mathcal{C}r^2-1\right]\left(e^{-\mathcal{C}r^2}-1\right).
\end{align}

\subsection{Physical Analysis of the Obtained Solutions}

The mass of spherically symmetric self-gravitating structure can be
calculated as
\begin{equation}\label{g63}
m(r)=\frac{1}{2}\int_{0}^{R}r^2 \hat{\mu}dr.
\end{equation}
By considering the initial condition $m(0)=0$ and solving the above
equation numerically, one can obtain the mass of spherical
anisotropic star. Another influential ingredient of self-gravitating
object is the compactness parameter $(\zeta(r))$ which can be
determined by taking the ratio of object mass and radius. The upper
limit of $\zeta(r)$ was constructed by Buchdahl \cite{42a} through
smooth matching of the interior spherical geometry and Schwarzschild
exterior solution. This limit is set
$\zeta(r)=\frac{m}{R}<\frac{4}{9}$ for a stable stellar
configuration, where $m(r)=\frac{R}{2}\left(1-e^{-\chi}\right)$. The
strong gravitational pull practiced by a celestial body increases
the wavelength of electromagnetic diffusion occurring in that body,
which is measured by a parameter known as redshift whose expression
is $D(r)=\frac{1}{\sqrt{1-2\zeta}}-1$. For a perfect fluid
configuration, Buchdahl restrained this parameter as $D(r)<2$ at the
surface of a star, whereas, its upper limit switches to 5.211 for
the case of anisotropic stellar object \cite{42b}.

The presence of normal matter as well as viability of the obtained
solutions can be checked by means of some constraints, known as
energy conditions. The parameters which govern the inner geometry of
the stellar configuration (composed of ordinary matter) obey these
conditions. These bounds are categorized as null, weak, strong and
dominant energy conditions. In
$f(\mathcal{R},\mathcal{T},\mathcal{Q})$ theory, these conditions
become
\begin{eqnarray}\nonumber
&&\hat{\mu}+\frac{s^2}{8\pi r^4} \geq 0, \quad \hat{\mu}+\hat{P}_{r}
\geq 0,\\\nonumber &&\hat{\mu}+\hat{P}_{\bot}+\frac{s^2}{4\pi r^4}
\geq 0, \quad \hat{\mu}-\hat{P}_{r}+\frac{s^2}{4\pi r^4} \geq
0,\\\label{g50} &&\hat{\mu}-\hat{P}_{\bot} \geq 0, \quad
\hat{\mu}+\hat{P}_{r}+2\hat{P}_{\bot}+\frac{s^2}{4\pi r^4} \geq 0.
\end{eqnarray}
In astrophysics, the stability of a star plays a crucial role in the
analysis of physically feasible system. Here, we use two criteria to
check the stability of resulting solutions. The first one is the
causality condition which states that the squared sound speed must
be within the range $[0,1]$, i.e., $0 \leq v_{s}^{2} \leq 1$. For
the case of anisotropic fluid, the discrepancy between squared sound
speeds in tangential
$(v_{s\bot}^{2}=\frac{d\hat{P}_{\bot}}{d\hat{\mu}})$ and radial
directions $(v_{sr}^{2}=\frac{d\hat{P}_{r}}{d\hat{\mu}})$ can be
employed to find the stable region of compact stars as $\mid
v_{s\bot}^{2}-v_{sr}^{2} \mid \leq 1$. Another factor which plays a
significant role in the analysis of stability of celestial objects
is adiabatic index $(\Gamma)$. Its value should not be less than
$\frac{4}{3}$ in the case of stable structures \cite{42c,43,44}.
Here, $\Gamma$ is defined as
\begin{equation}\label{g62}
\hat{\Gamma}=\frac{\hat{\mu}+\hat{P}_{r}}{\hat{P}_{r}}
\left(\frac{d\hat{P}_{r}}{d\hat{\mu}}\right).
\end{equation}
\begin{figure}\center
\epsfig{file=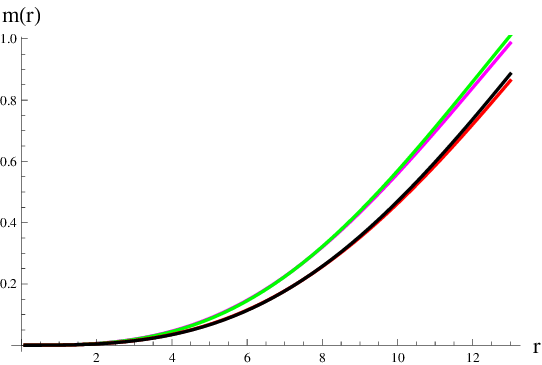,width=0.4\linewidth}\epsfig{file=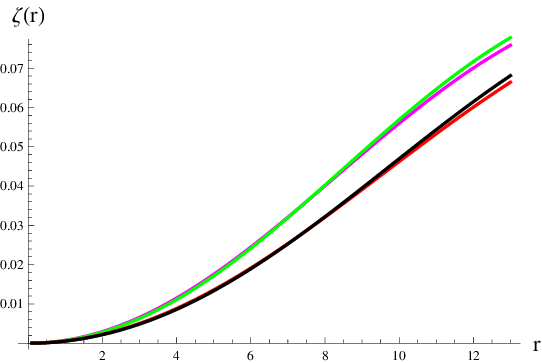,width=0.4\linewidth}
\epsfig{file=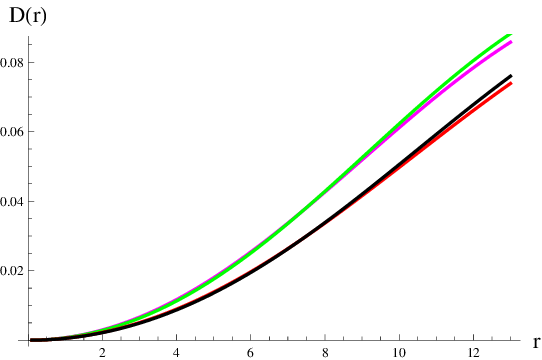,width=0.4\linewidth} \caption{Plots of
mass, compactness and redshift parameters corresponding to
$S_{0}=0.1,~\varphi=0.1$ (pink),~$\varphi=0.9$ (green) and
$S_{0}=0.8,~\varphi=0.1$ (red),~$\varphi=0.9$ (black) for
solution-I.}
\end{figure}

We take a linear model \cite{22} to examine the physical feasibility
as well as stability of the obtained solutions as
\begin{equation}\label{g61}
f(\mathcal{R},\mathcal{T},\mathcal{Q})=\mathcal{R}+\beta
\mathcal{Q},
\end{equation}
where $\beta$ is an arbitrary constant. To check physical behavior
of the solution-I, we take $\beta=-0.1$ and set $\mathcal{C}$ as a
constant given in Eq.\eqref{g43}. The values of remaining two
parameters $\mathcal{A}$ and $\mathcal{B}$ are presented in
Eqs.\eqref{g37} and \eqref{g37a}. Figure \textbf{1} (left) exhibits
the mass of spherical geometry \eqref{g6} for different values of
decoupling parameter $\varphi$ and charge. It can be seen that the
mass increases with the increase of $\varphi$ whereas it decreases
as charge increases. The compactness and redshift parameters are
shown in Figure \textbf{1} (right and lower) whose ranges are found
to be in accordance with their respective limits.
\begin{figure}\center
\epsfig{file=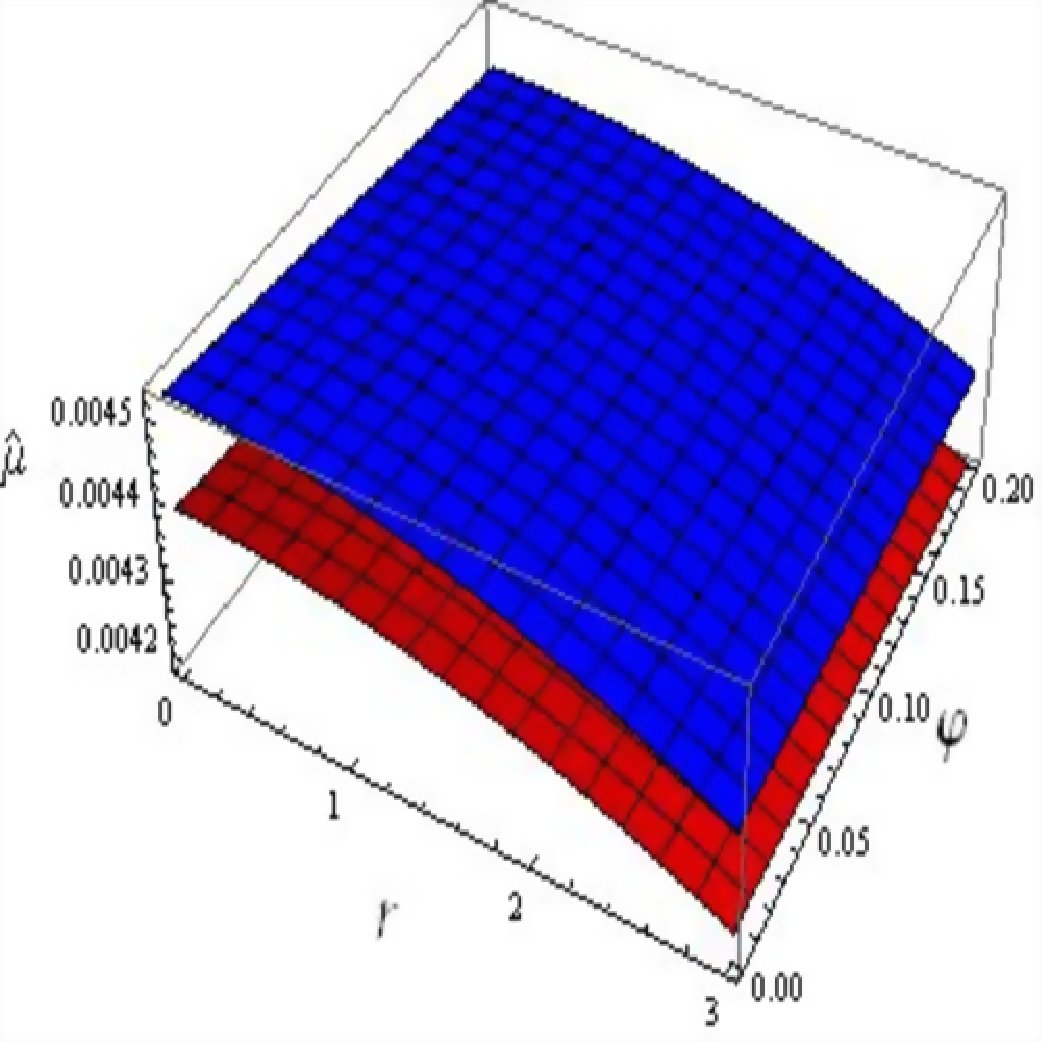,width=0.4\linewidth}\epsfig{file=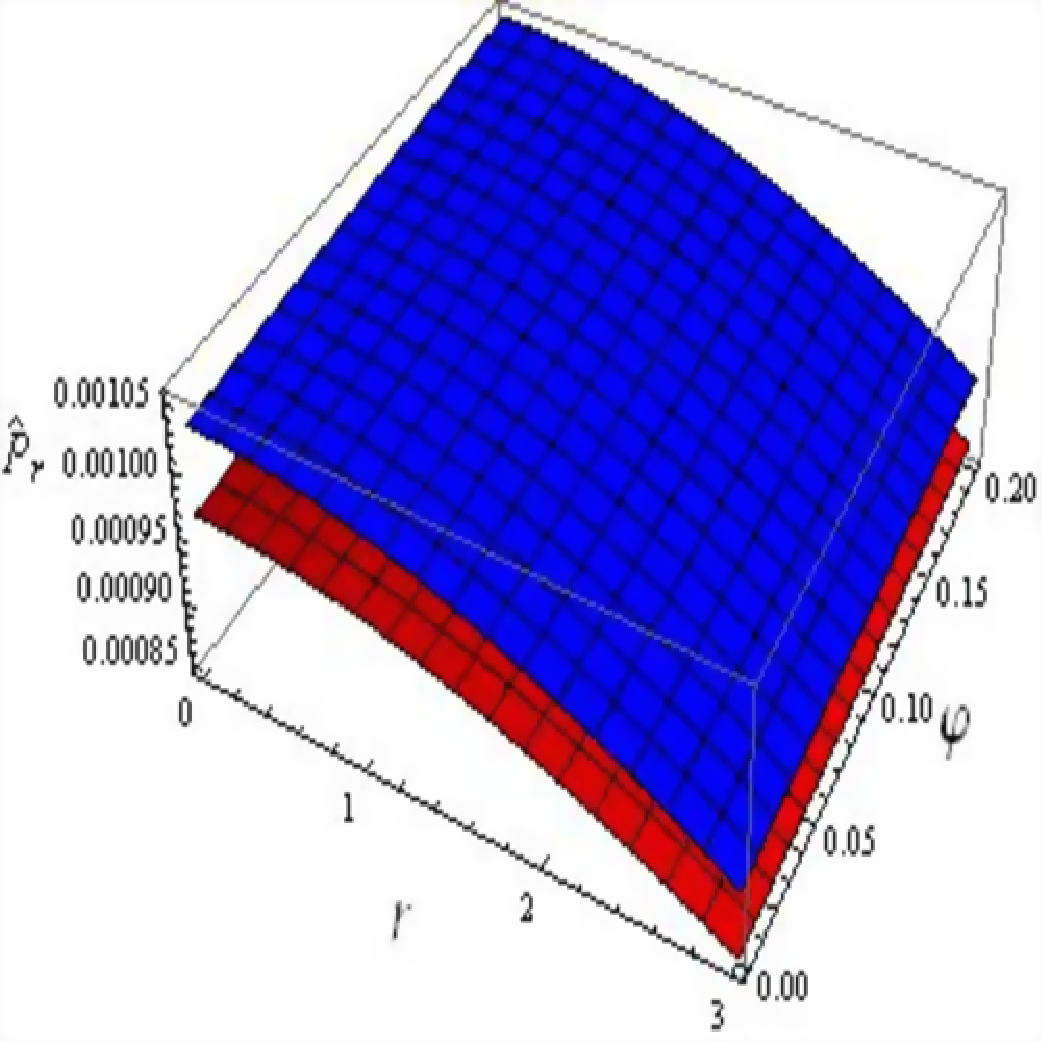,width=0.4\linewidth}
\epsfig{file=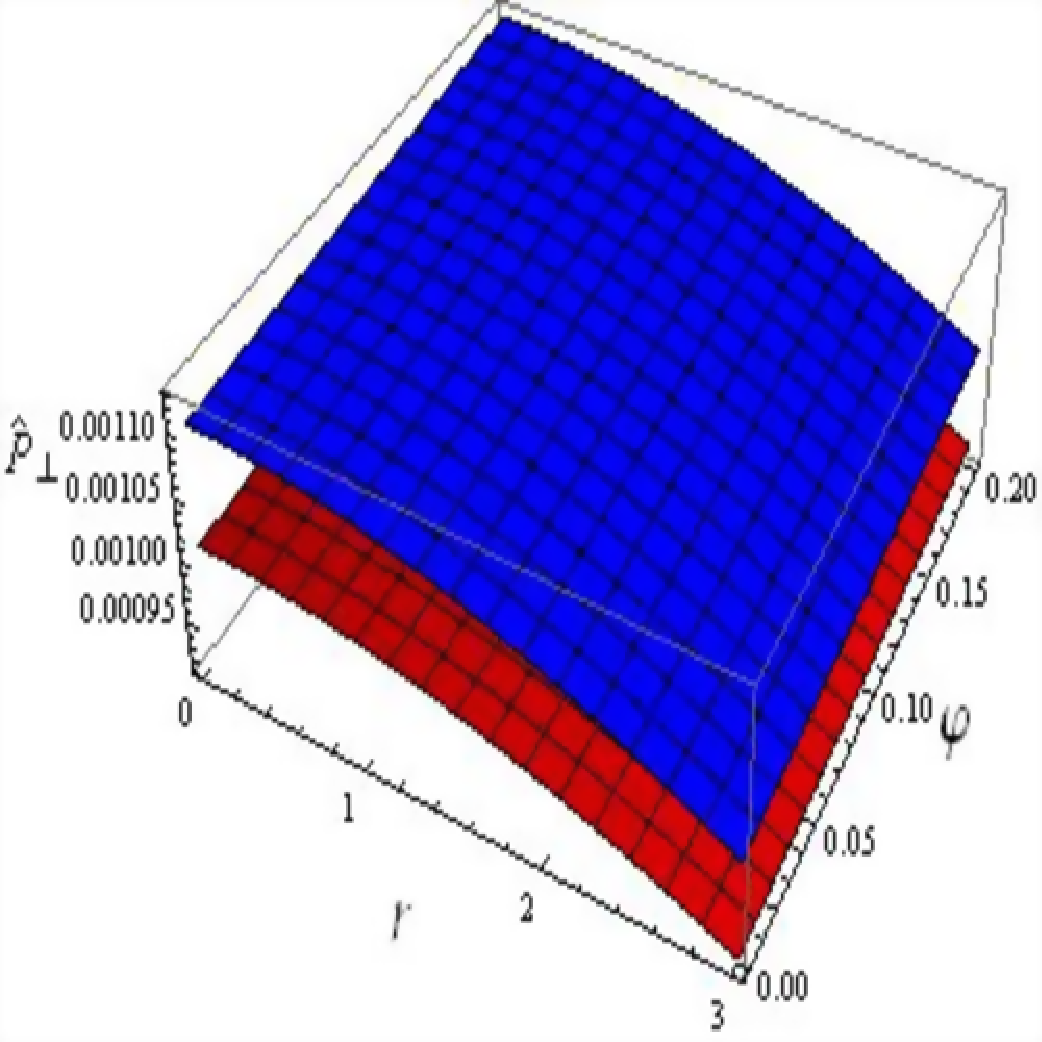,width=0.4\linewidth}\epsfig{file=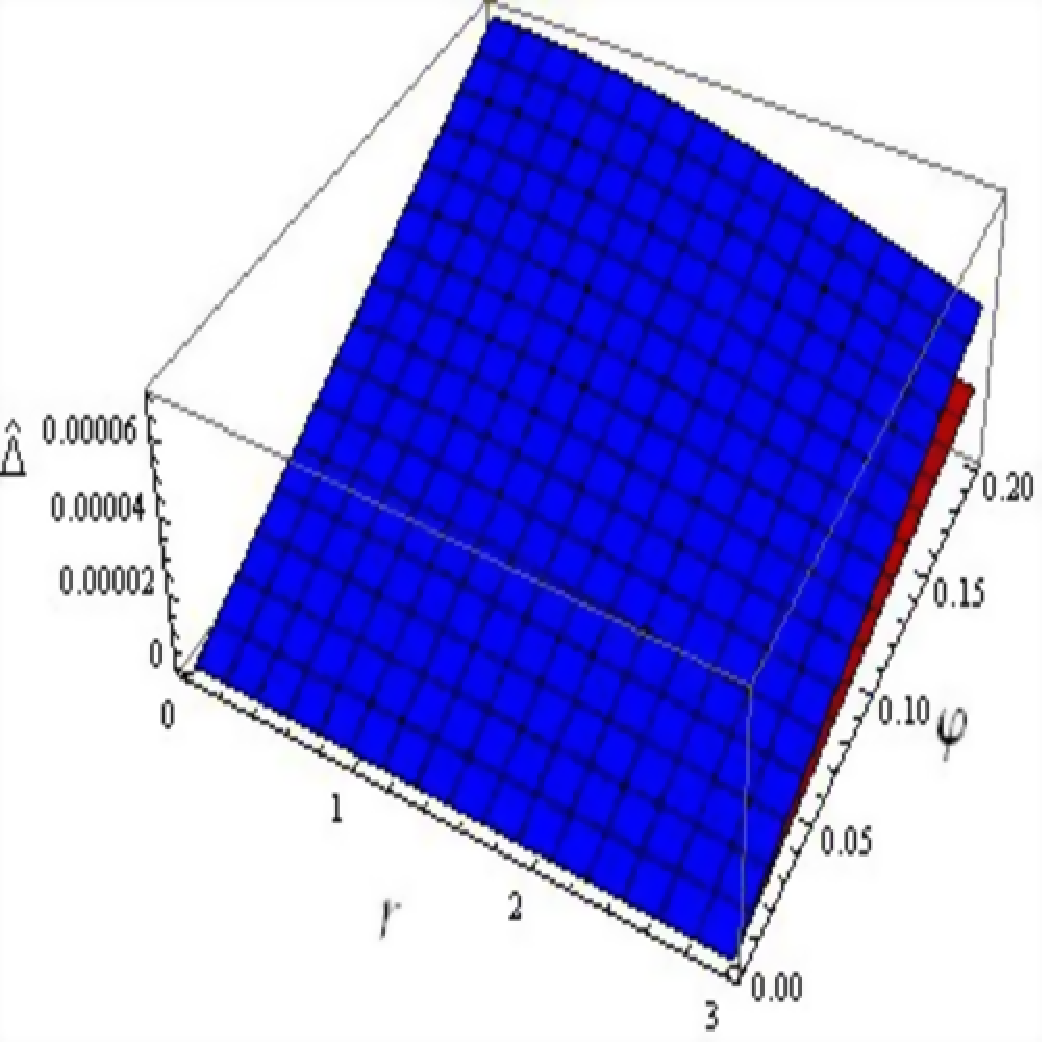,width=0.4\linewidth}
\caption{Plots of $\hat{\mu},~\hat{P}_{r},~\hat{P}_{\bot}$ and
$\hat{\Delta}$ versus $r$ and $\varphi$ with $S_{0}=0.1$ (Blue),
$S_{0}=0.8$ (Red), $M_{0}=1M_{\bigodot}$ and
$R=(0.2)^{-1}M_{\bigodot}$ for solution-I.}
\end{figure}
\begin{figure}\center
\epsfig{file=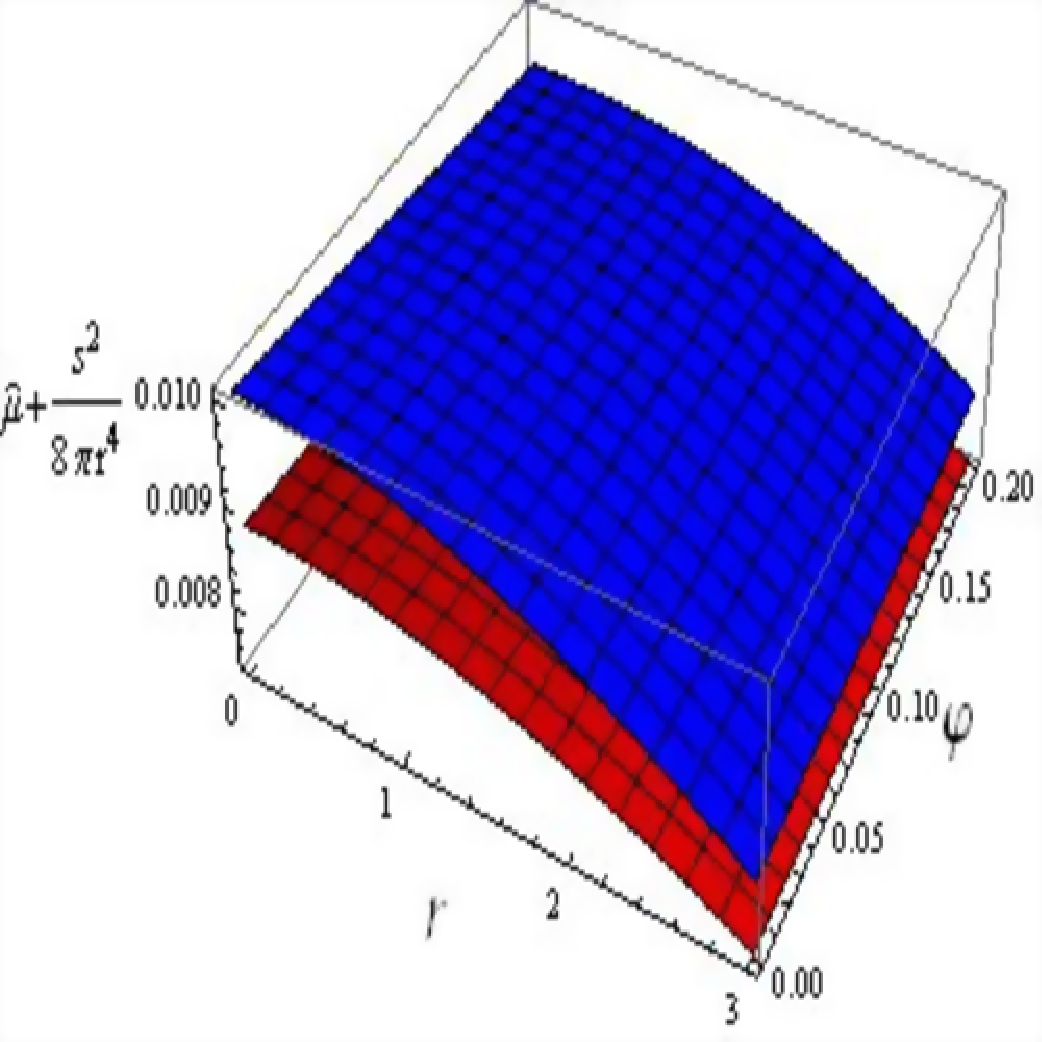,width=0.4\linewidth}\epsfig{file=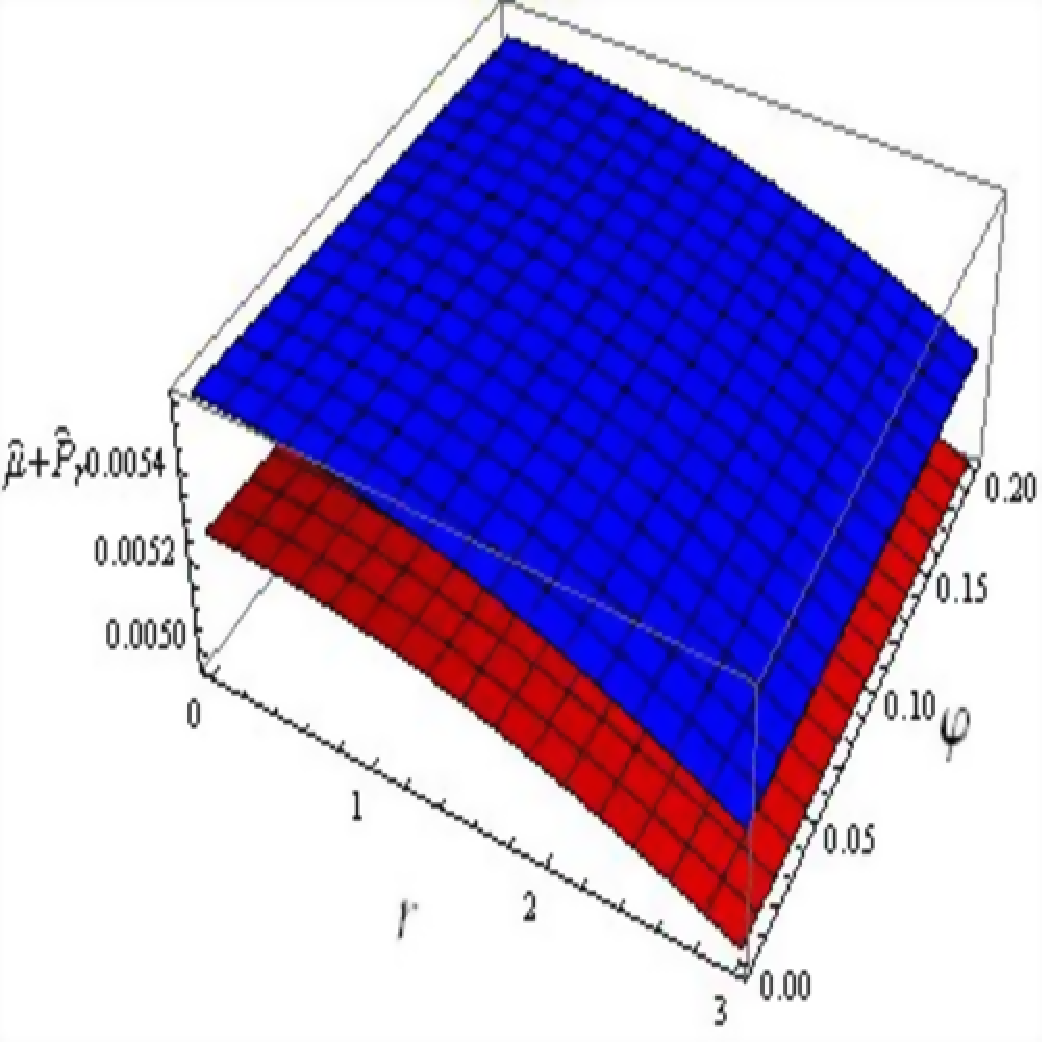,width=0.4\linewidth}
\epsfig{file=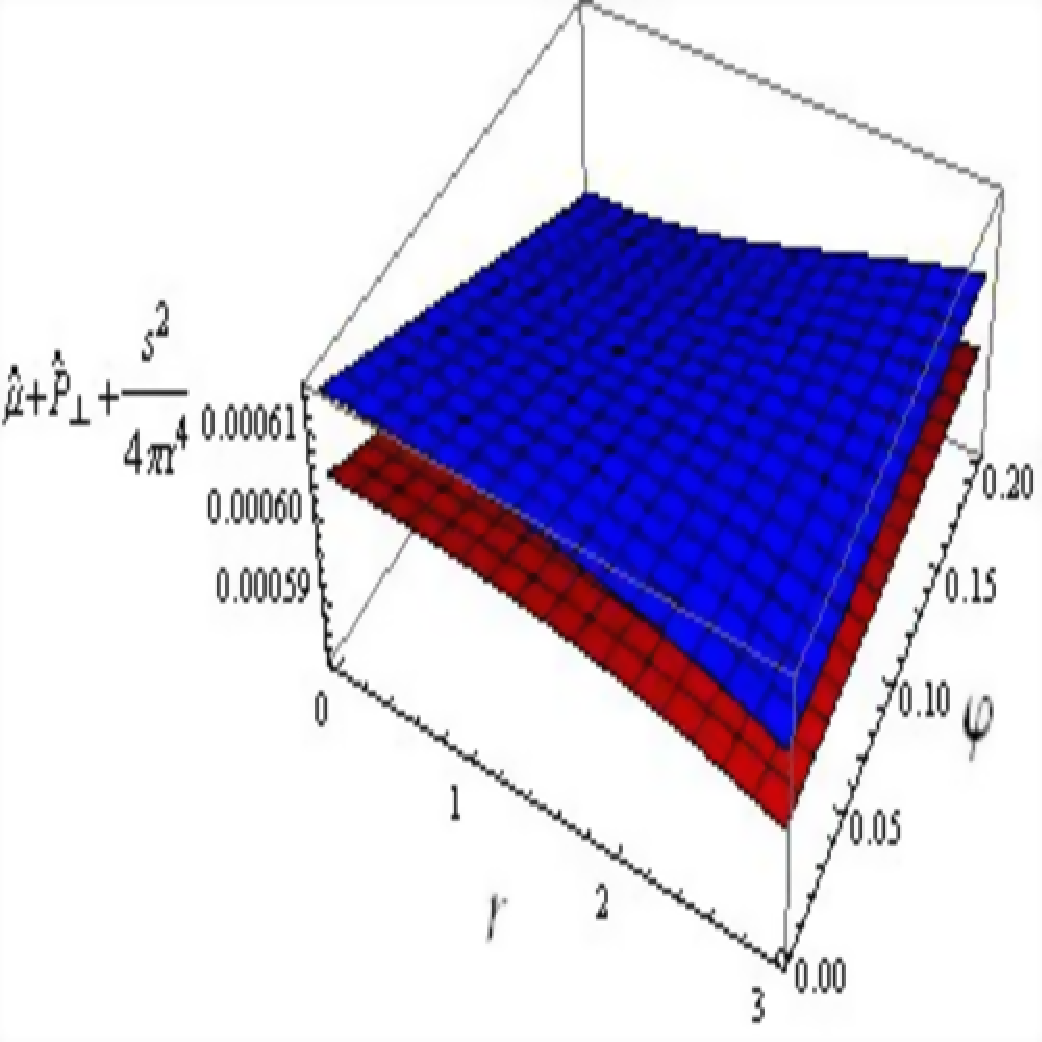,width=0.4\linewidth}\epsfig{file=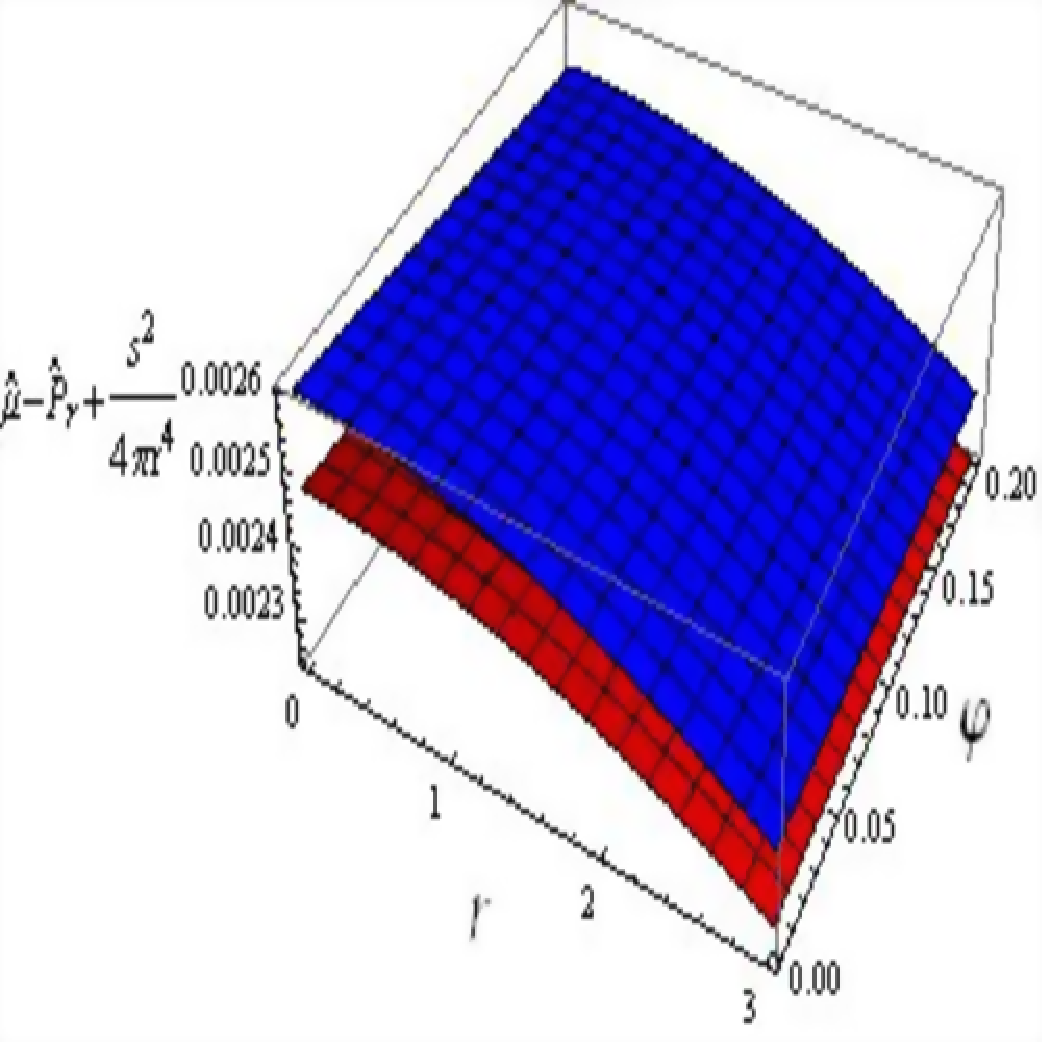,width=0.4\linewidth}
\epsfig{file=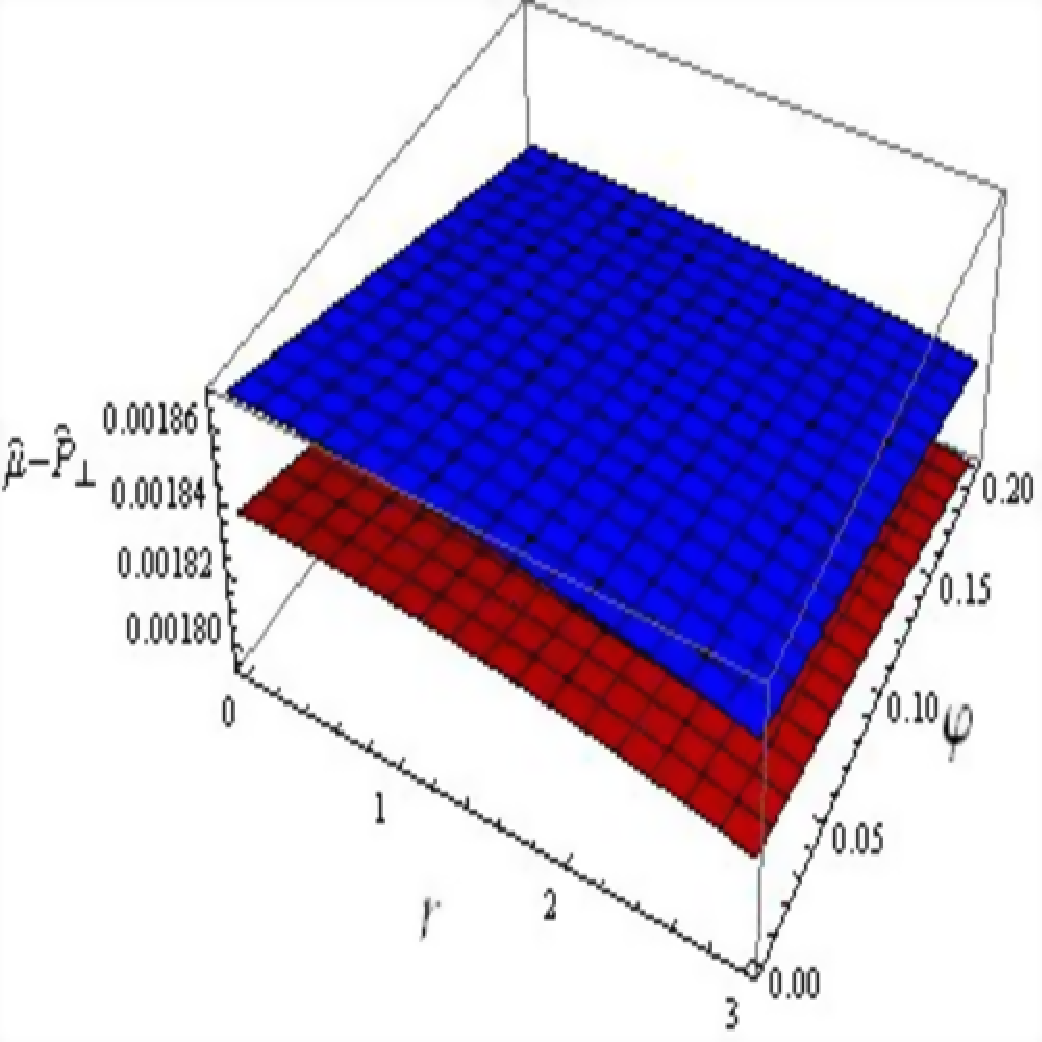,width=0.4\linewidth}\epsfig{file=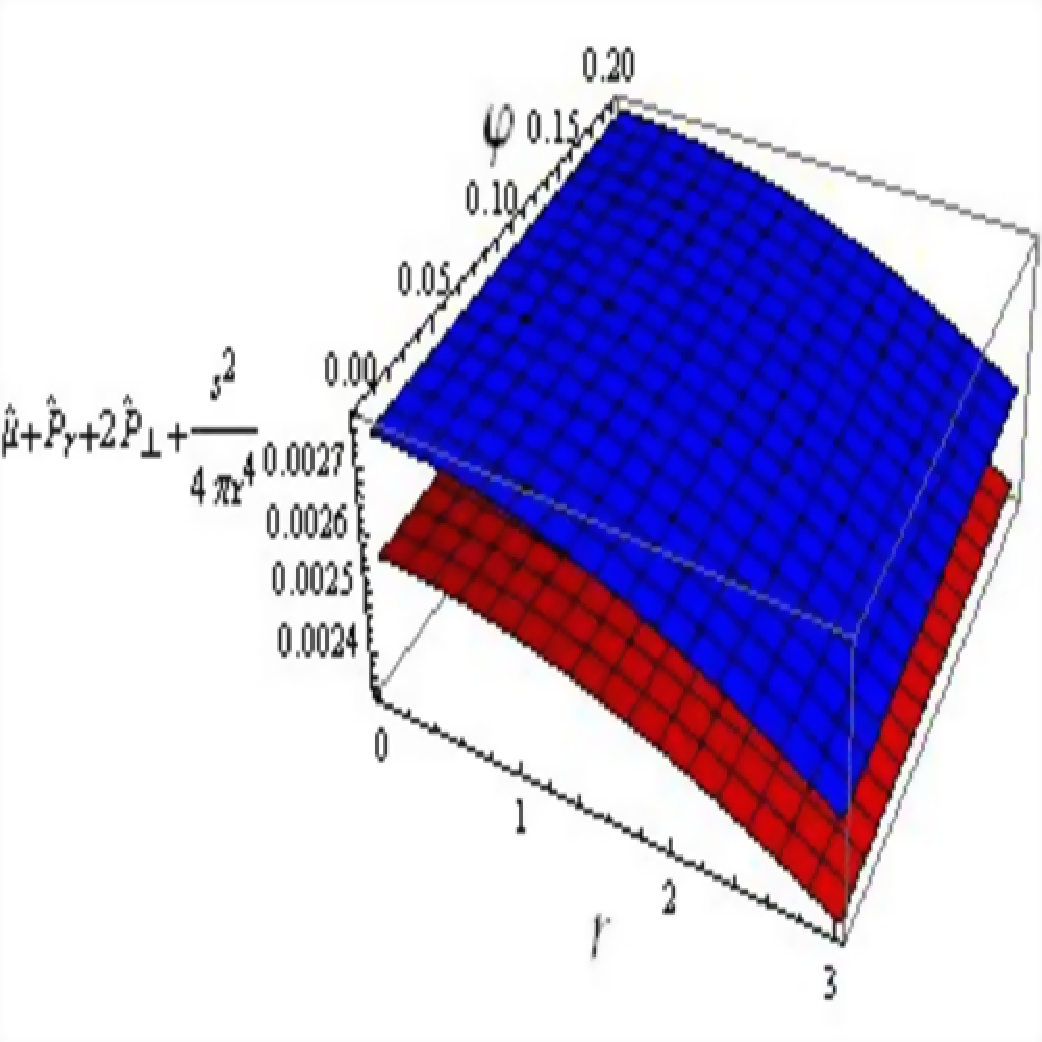,width=0.4\linewidth}
\caption{Plots of energy conditions versus $r$ and $\varphi$ with
$S_{0}=0.1$ (Blue), $S_{0}=0.8$ (Red), $M_{0}=1M_{\bigodot}$ and
$R=(0.2)^{-1}M_{\bigodot}$ for solution-I.}
\end{figure}

The values of matter variables (such as energy density, radial and
tangential pressures) should be positive, finite and maximum at the
center of an astrophysical object, whereas their behavior must be
monotonically decreasing towards the star's boundary. Figure
\textbf{2} (upper left) indicates that the effective energy density
in the middle is maximum but decreases as $r$ increases. It can also
be observed that the density decreases linearly with the increase of
decoupling parameter $\varphi$ as well as charge. Thus, more the
charge is, less dense is the sphere. The graphs of $\hat{P}_{r}$ and
$\hat{P}_{\bot}$ are found to be in a similar fashion for the
parameter $\beta$ and decrease by increasing $r$ and charge
parameter. Also, both ingredients show increasing behavior with rise
in $\varphi$. Figure \textbf{2} (lower right) shows that anisotropic
factor $\hat{\Delta}$ is zero at the middle for different values of
charge while increases with the increase in $\varphi$ which confirms
that the decoupling parameter $\varphi$ produces stronger anisotropy
in the system. Figure \textbf{3} demonstrates the physical viability
of our established anisotropic solution-I as it fulfills all energy
conditions \eqref{g50}. Both plots of Figure \textbf{4} indicate
that the solution-I meets stability requirements for higher values
of the decoupling parameter $\varphi$. Figure \textbf{4} (left) also
exhibits that the stability of this solution enhances with increase
in charge.
\begin{figure}\center
\epsfig{file=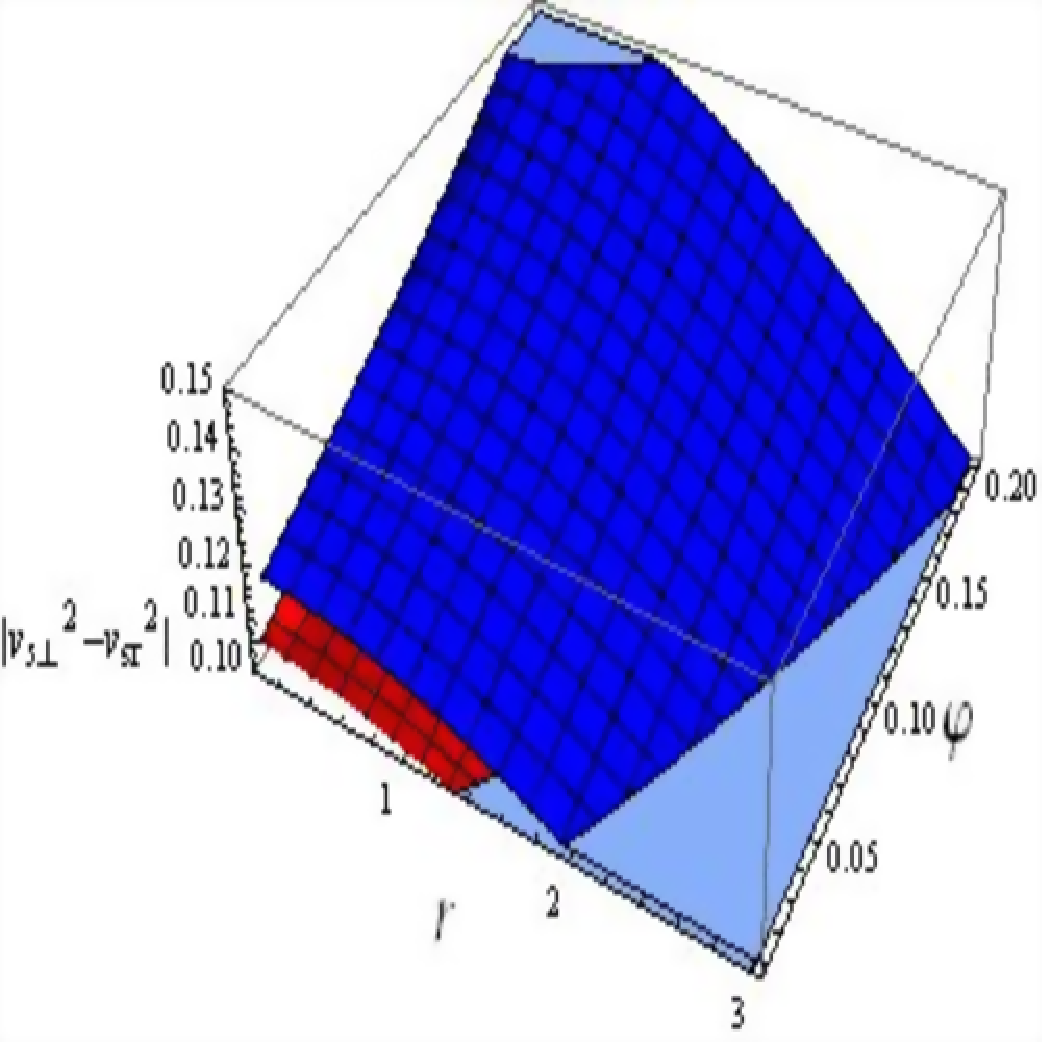,width=0.4\linewidth}\epsfig{file=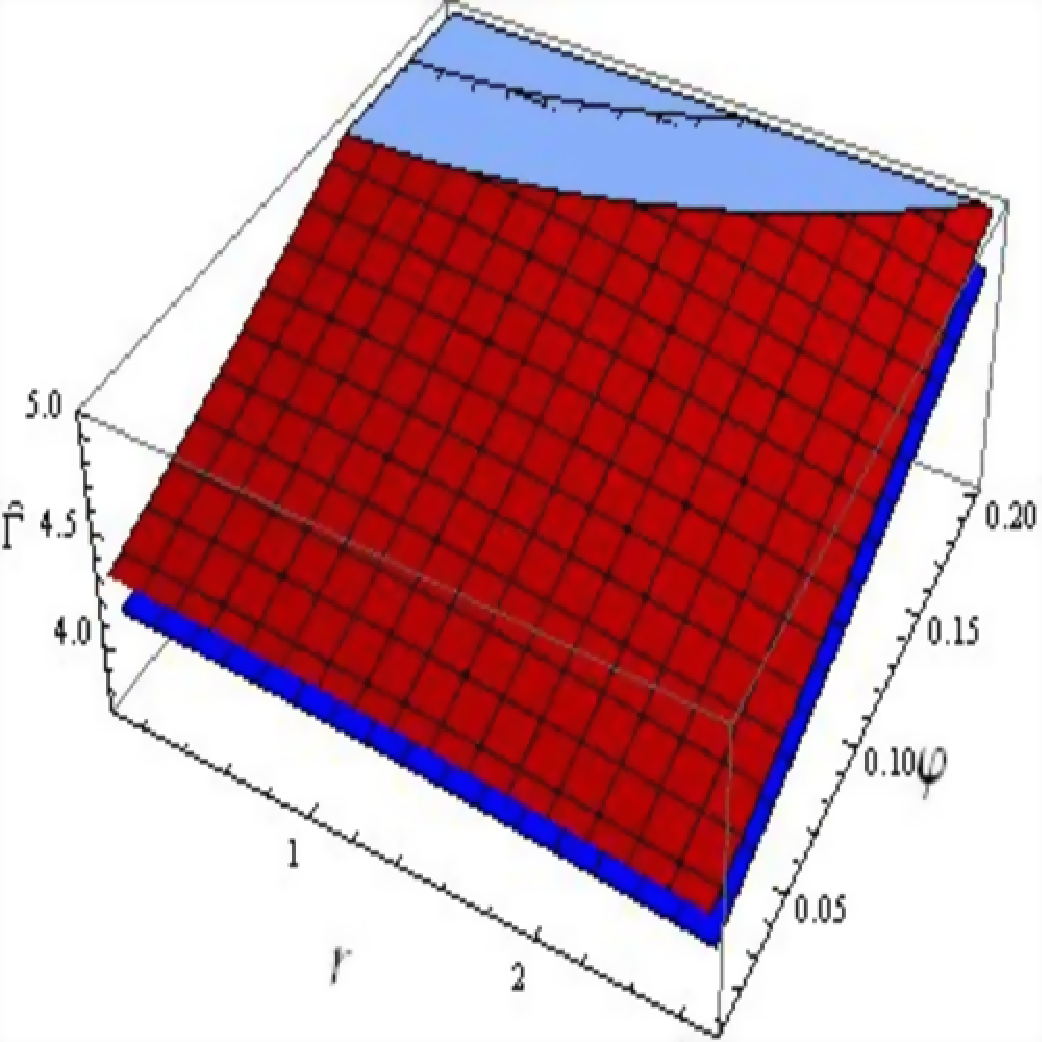,width=0.4\linewidth}
\caption{Plots of $|v_{s\bot}^2-v_{sr}^2|$ and adiabatic index
versus $r$ and $\varphi$ with $S_{0}=0.1$ (Blue), $S_{0}=0.8$ (Red),
$M_{0}=1M_{\bigodot}$ and $R=(0.2)^{-1}M_{\bigodot}$ for
solution-I.}
\end{figure}

We now investigate physical characteristics of the second solution
by taking $\beta=-0.05$. The values of parameters $\mathcal{C}$ and
$\mathcal{A}$ are shown in Eqs.\eqref{g38} and \eqref{g56}. Figure
\textbf{5} (upper left) represents that the mass of spherical body
\eqref{g6} decreases with rise in $\varphi$ as well as charge.
Figure \textbf{5} (upper right and lower) also indicates that the
compactness $(\zeta(r))$ and redshift $(D(r))$ fulfil the required
limits. Figure \textbf{6} illustrates the physical behavior of
$\hat{\mu},~\hat{P}_{r},~\hat{P}_{\bot}$ and $\hat{\Delta}$. By
increasing the parameter $\varphi$, the energy density also
increases whereas it decreases with increase in charge. The
effective pressure components in radial and tangential directions
show decreasing behavior with increase in both decoupling parameter
and charge. Figure \textbf{6} (lower right) reveals the increasing
behavior of anisotropy $(\hat{\Delta})$ towards surface. The
solution-II meets all energy bounds \eqref{g50}, thus it is
physically viable as can be seen from Figure \textbf{7}. Both plots
of Figure \textbf{8} confirm that our second solution is also stable
throughout the system. Figure \textbf{8} (left) shows that the
increment in charge causes the system less stable.
\begin{figure}\center
\epsfig{file=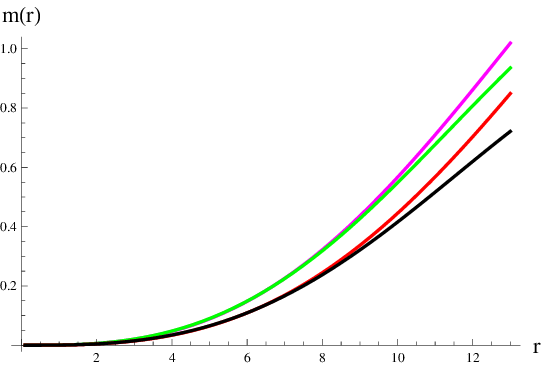,width=0.4\linewidth}\epsfig{file=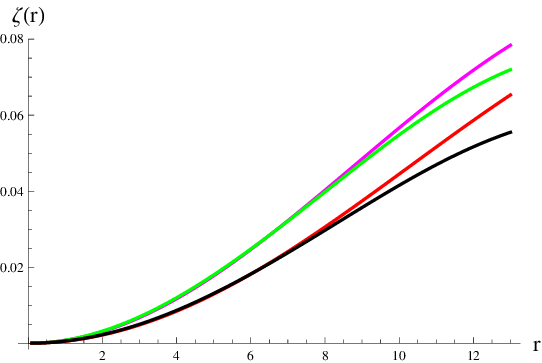,width=0.4\linewidth}
\epsfig{file=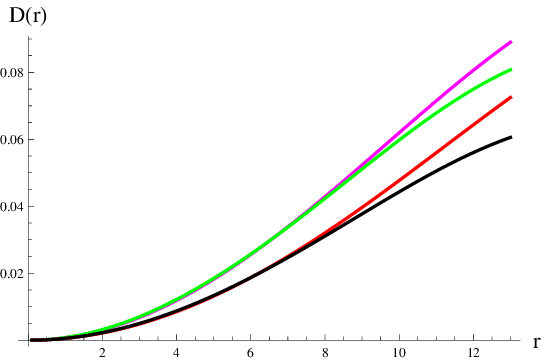,width=0.5\linewidth} \caption{Plots of
mass, compactness and redshift parameters corresponding to
$S_{0}=0.1,~\varphi=0.1$ (pink),~$\varphi=0.3$ (green) and
$S_{0}=0.8,~\varphi=0.1$ (red),~$\varphi=0.3$ (black) for
solution-II.}
\end{figure}
\begin{figure}\center
\epsfig{file=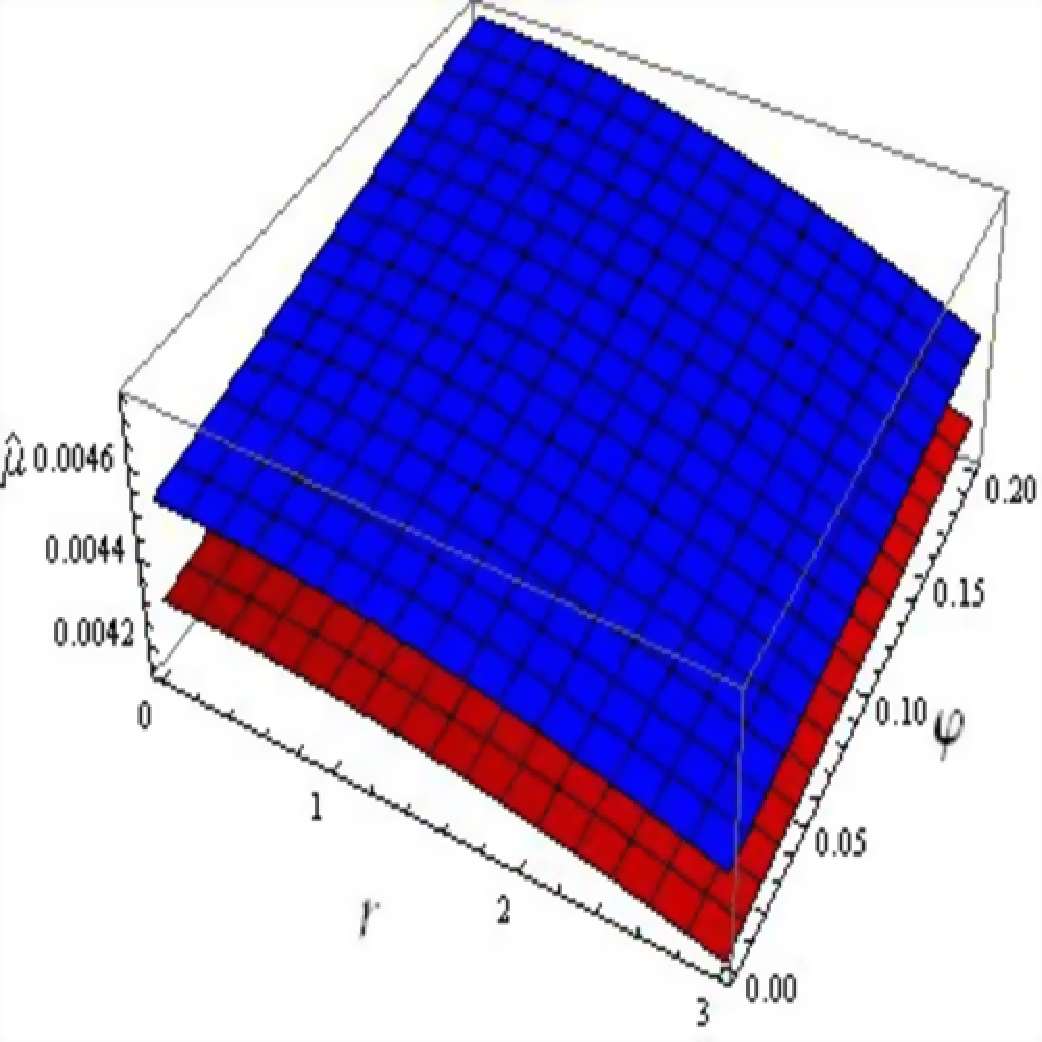,width=0.4\linewidth}\epsfig{file=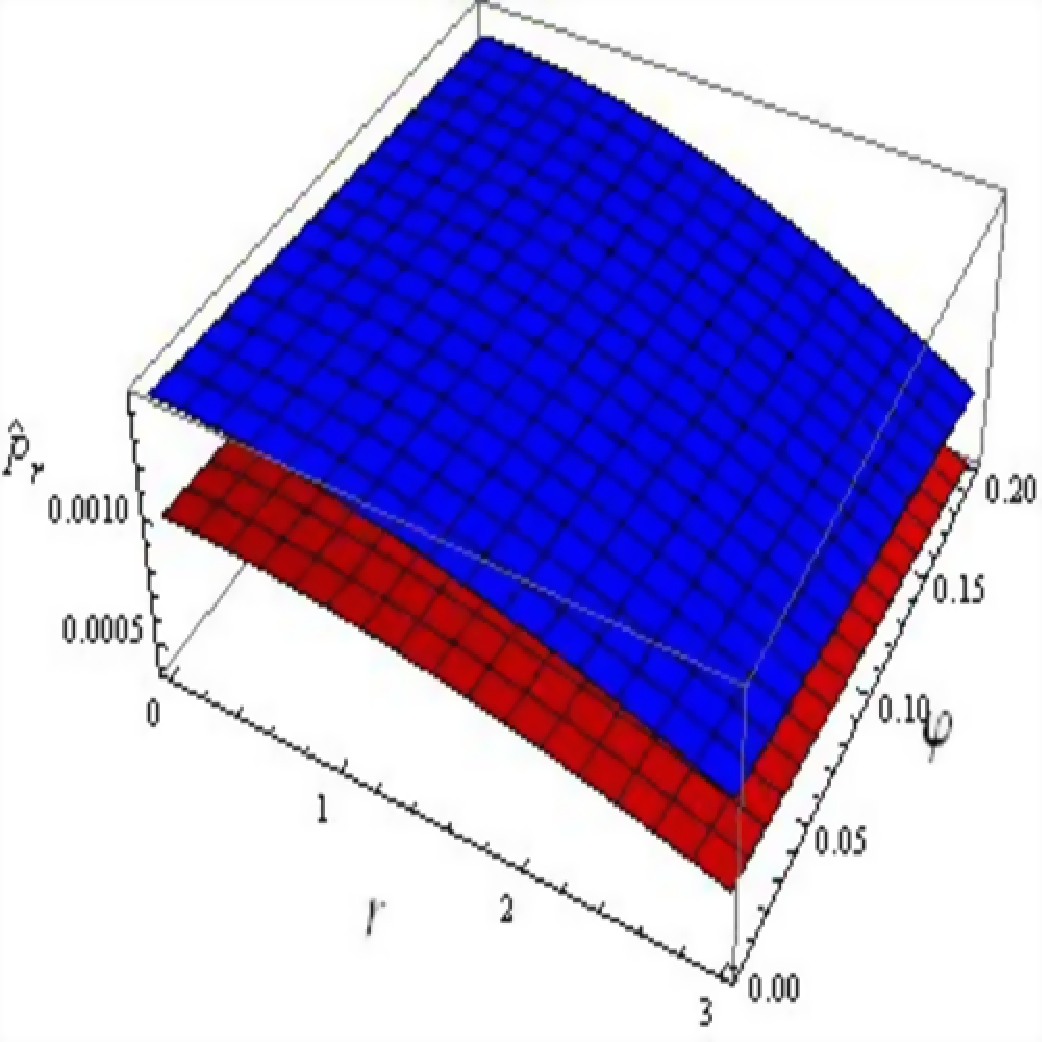,width=0.4\linewidth}
\epsfig{file=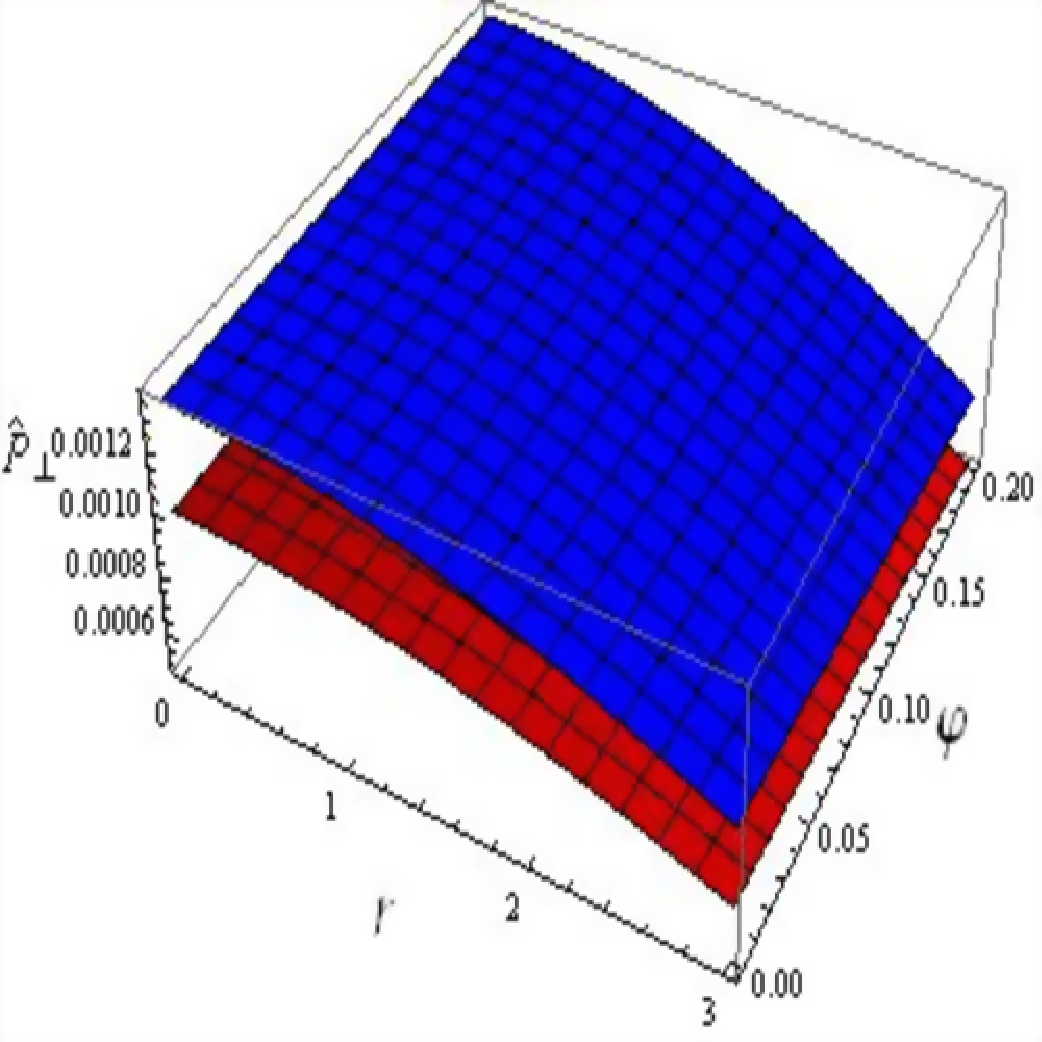,width=0.4\linewidth}\epsfig{file=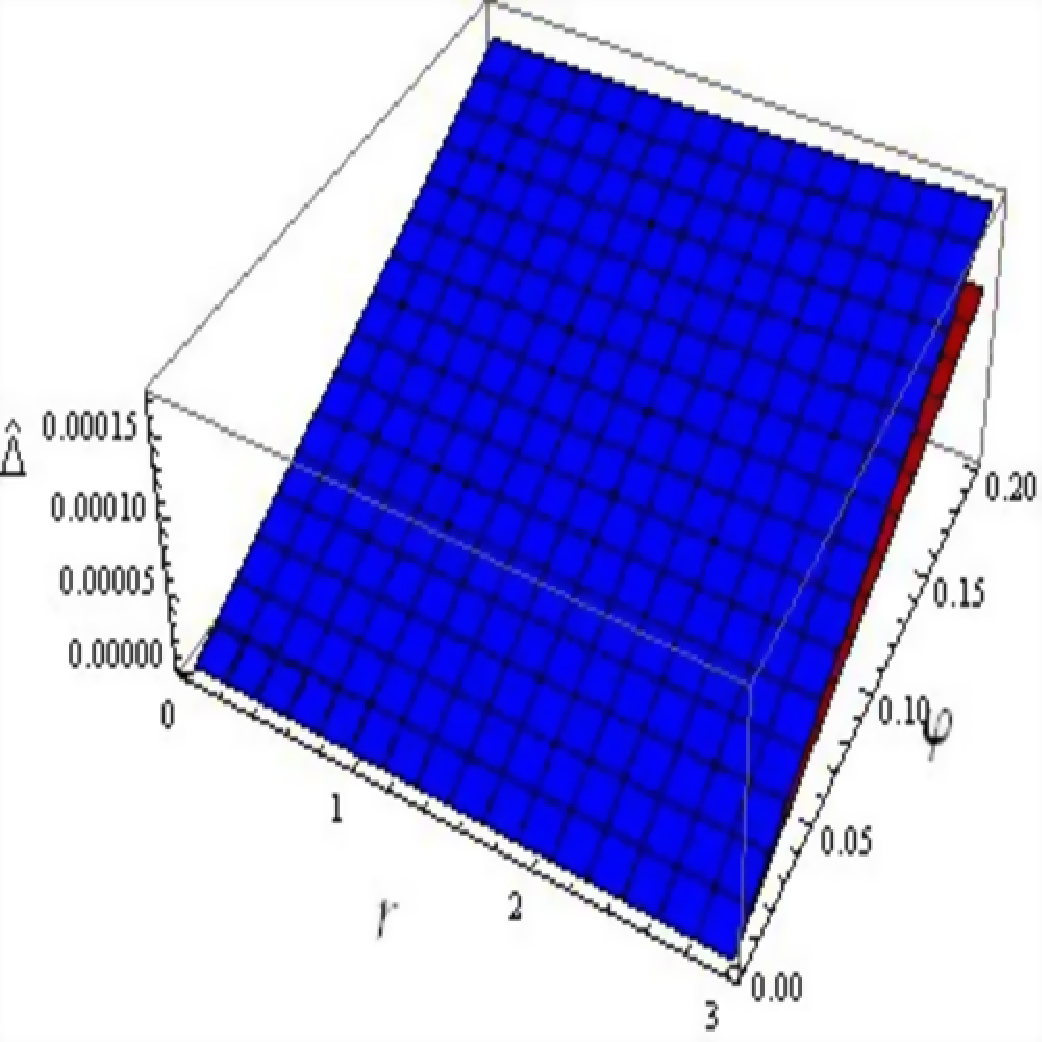,width=0.4\linewidth}
\caption{Plots of $\hat{\mu},~\hat{P}_{r},~\hat{P}_{\bot}$ and
$\hat{\Delta}$ versus $r$ and $\varphi$ with $S_{0}=0.1$ (Blue),
$S_{0}=0.8$ (Red), $M_{0}=1M_{\bigodot}$ and
$R=(0.2)^{-1}M_{\bigodot}$ for solution-II.}
\end{figure}
\begin{figure}\center
\epsfig{file=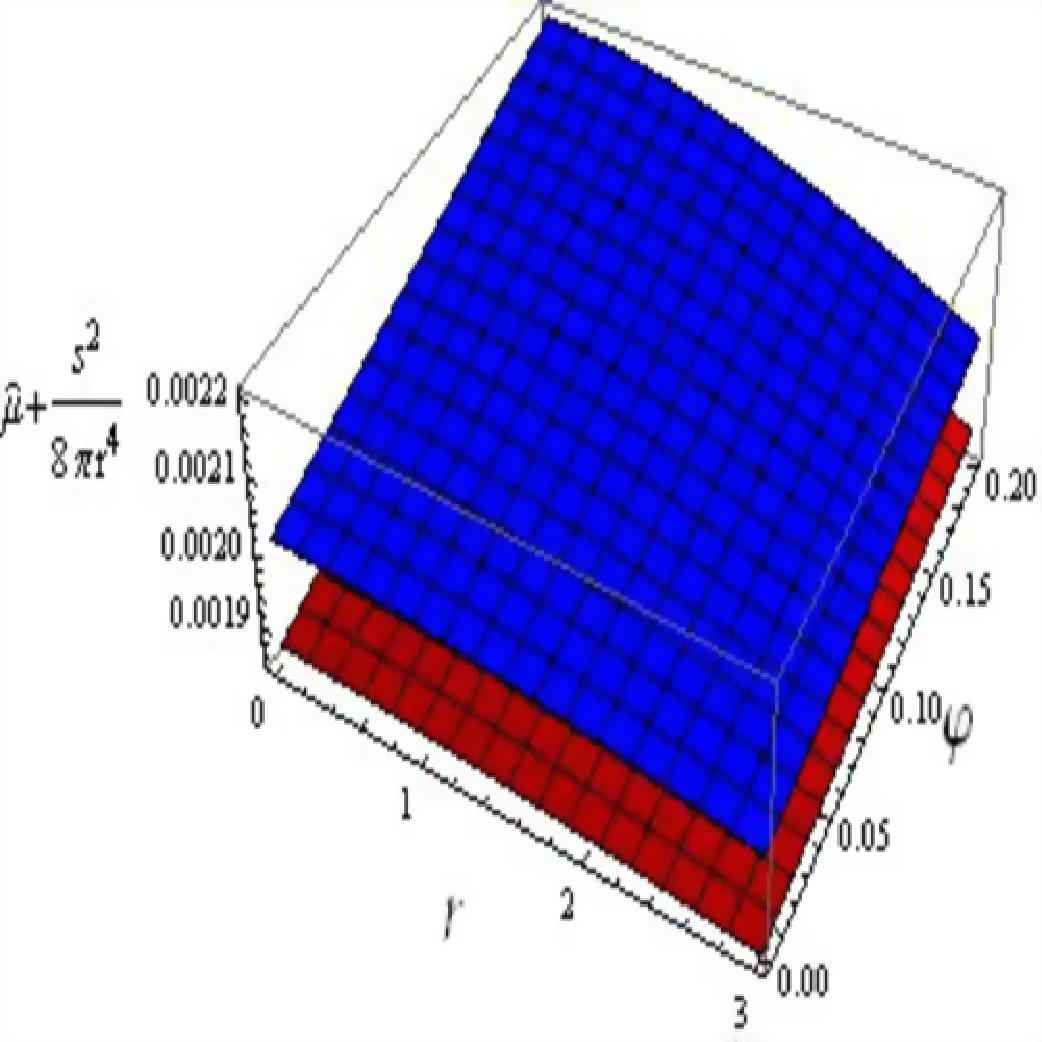,width=0.4\linewidth}\epsfig{file=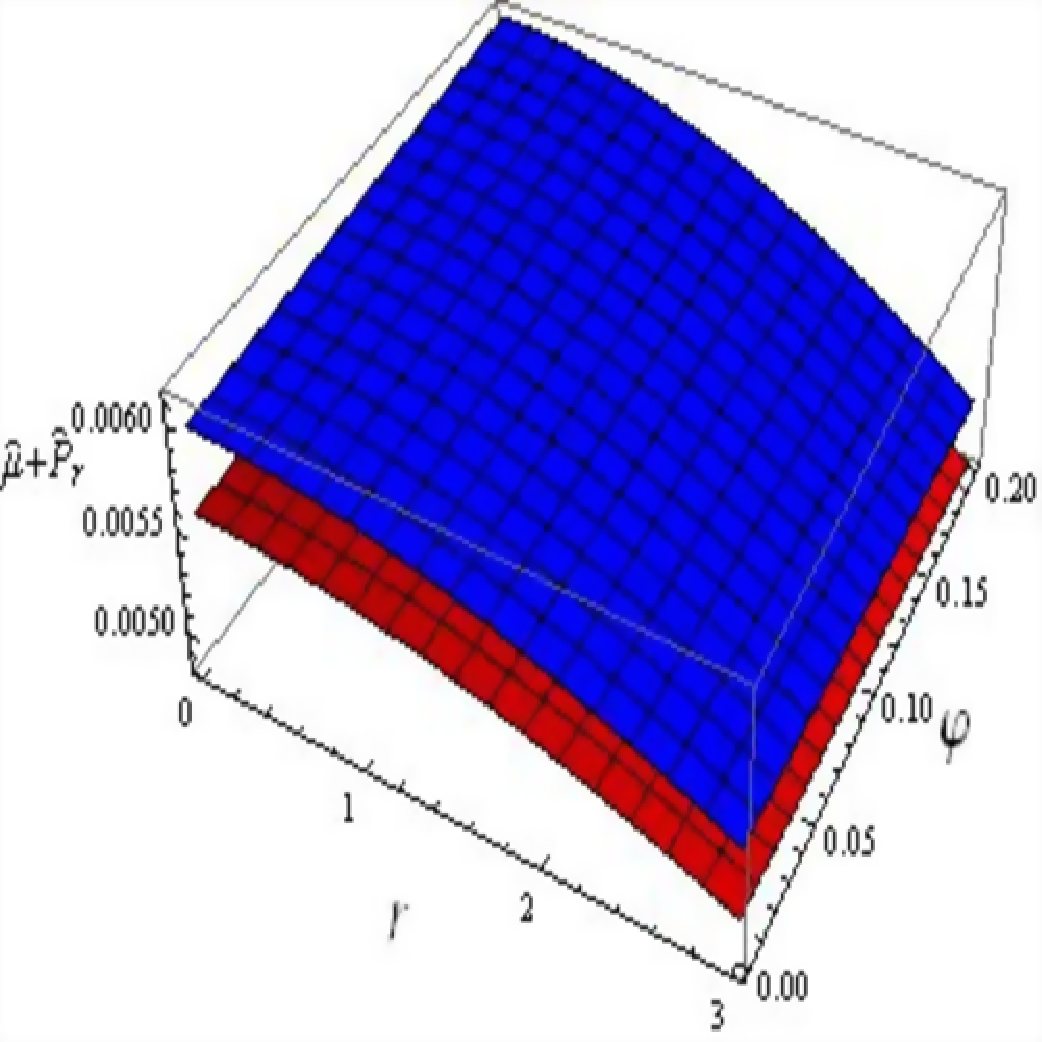,width=0.4\linewidth}
\epsfig{file=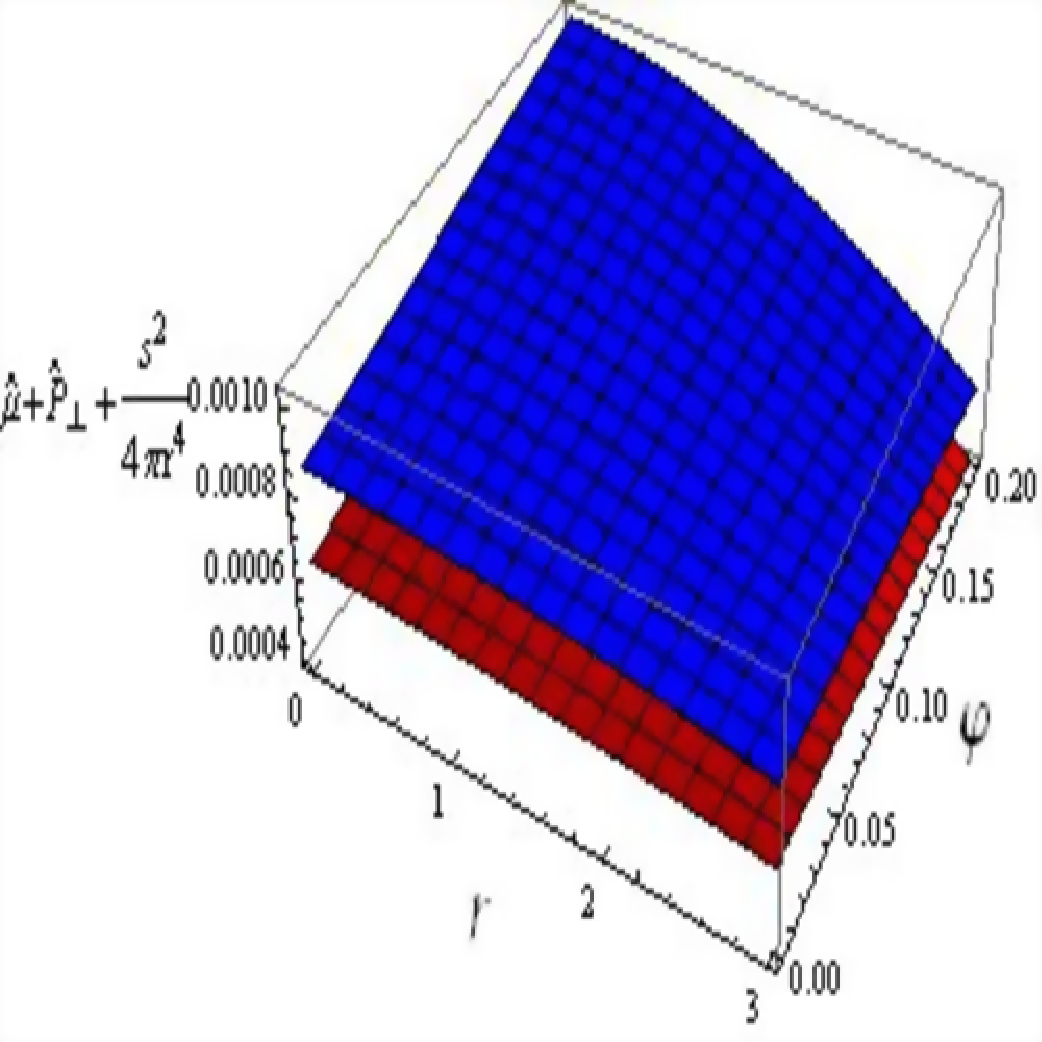,width=0.4\linewidth}\epsfig{file=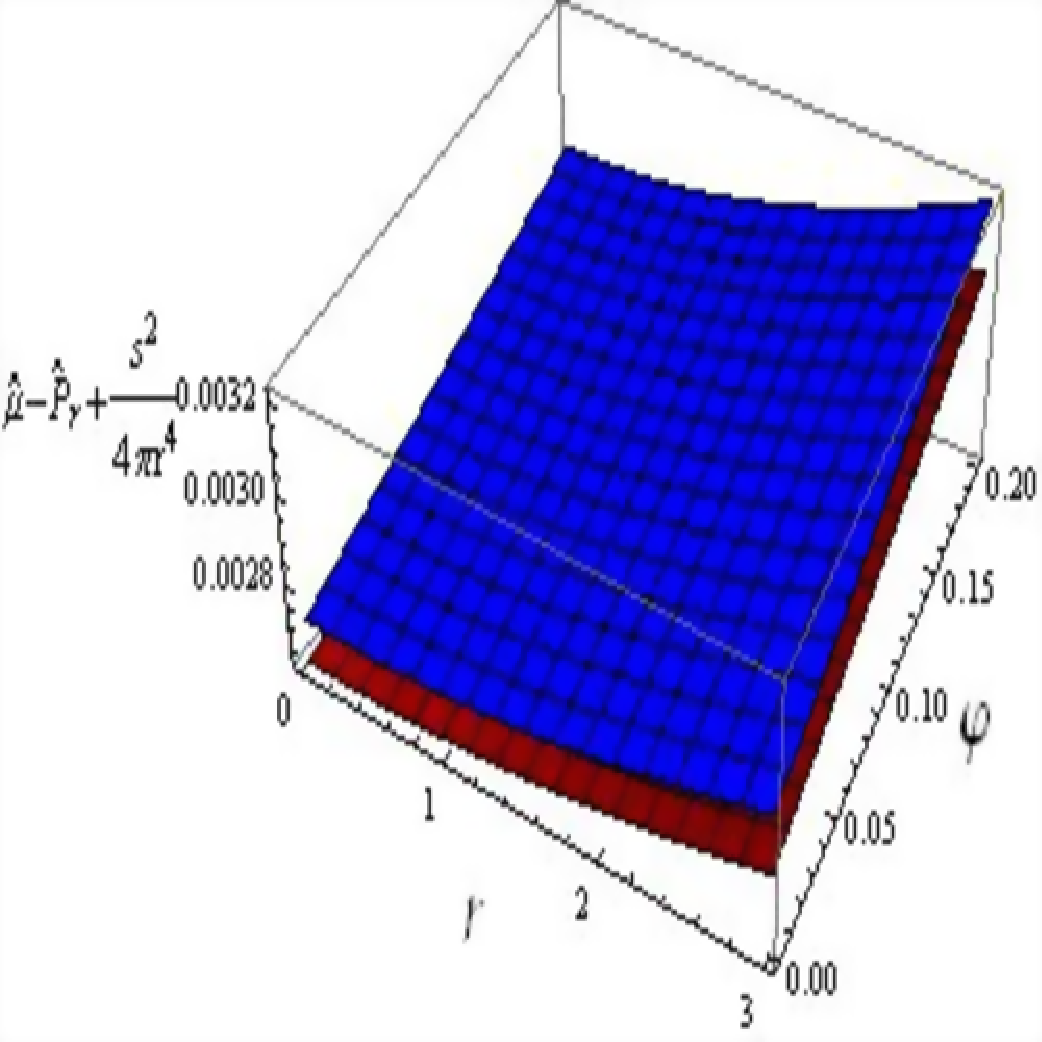,width=0.4\linewidth}
\epsfig{file=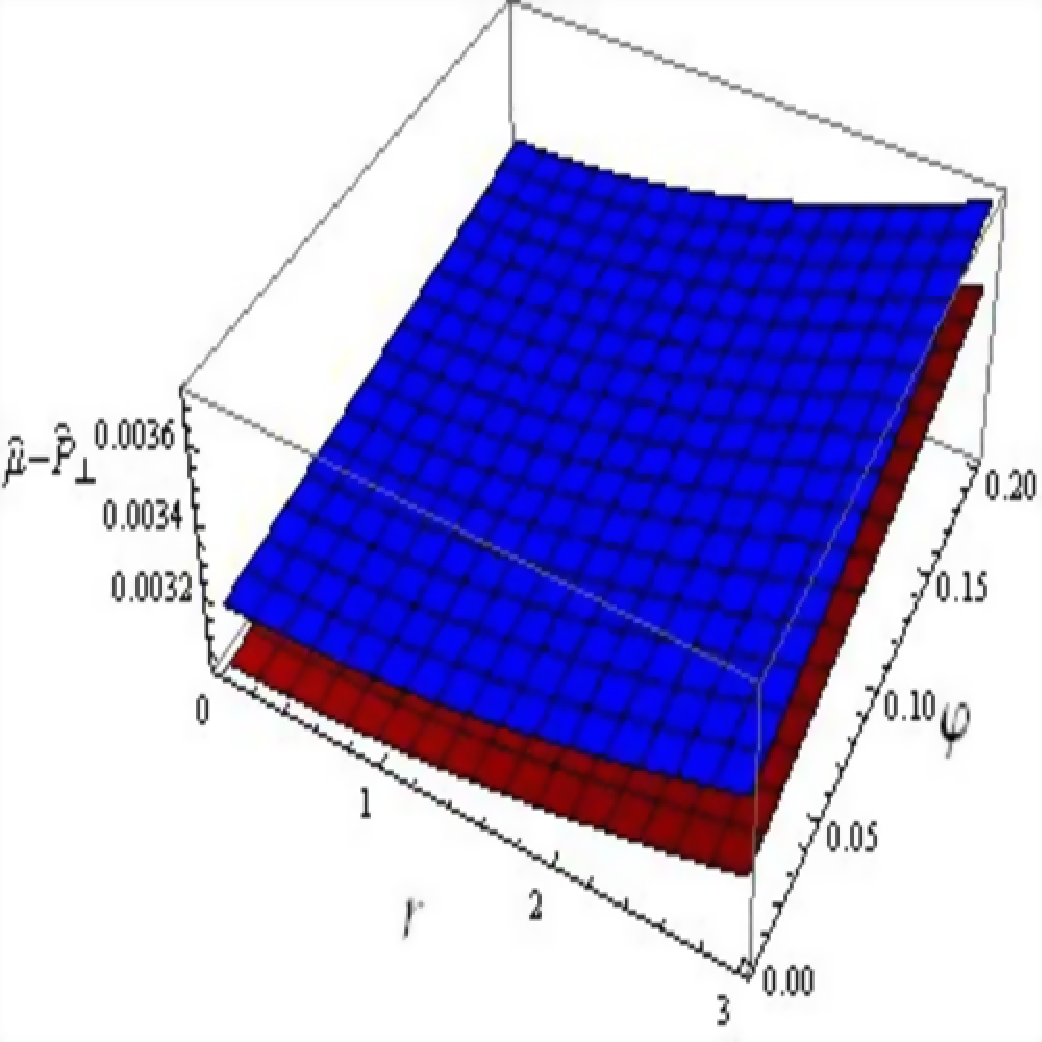,width=0.4\linewidth}\epsfig{file=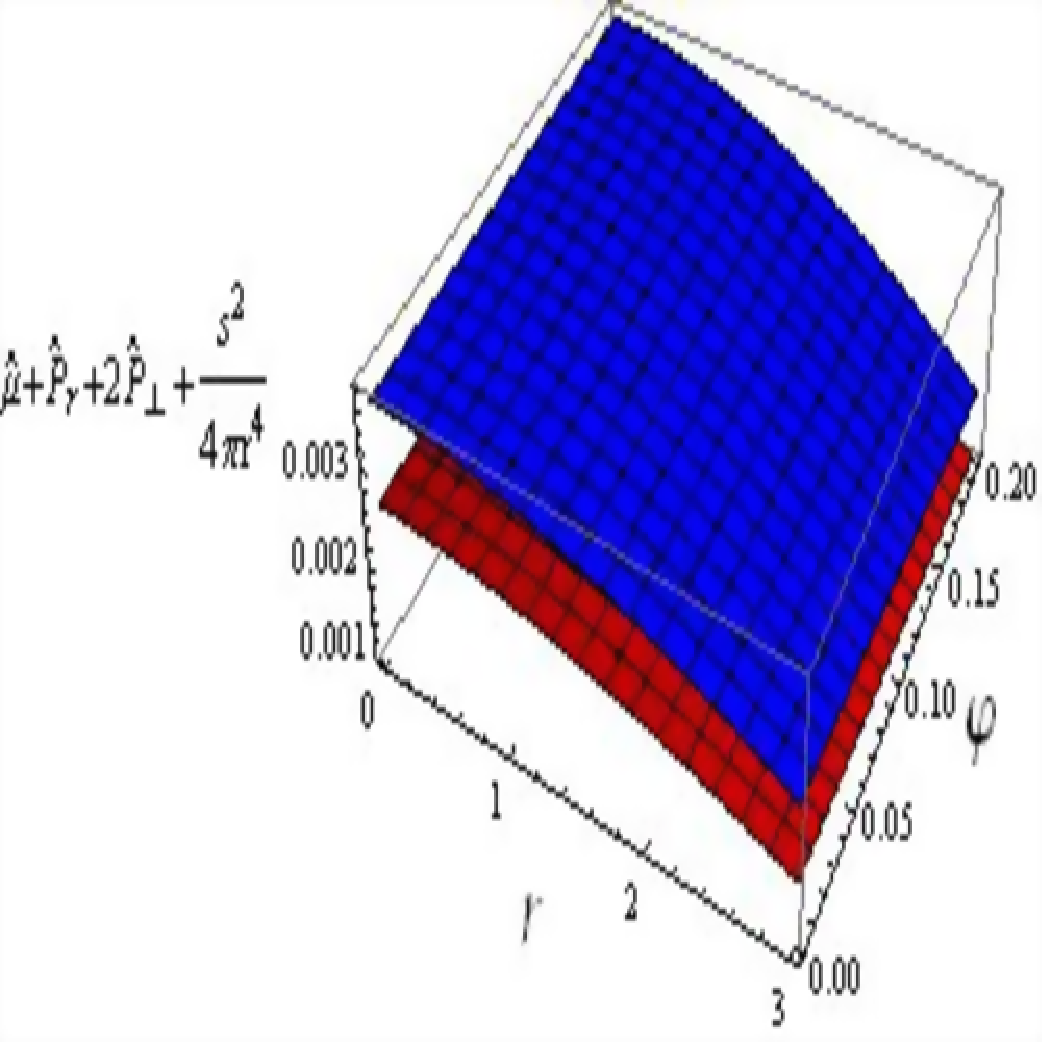,width=0.4\linewidth}
\caption{Plots of energy conditions versus $r$ and $\varphi$ with
$S_{0}=0.1$ (Blue), $S_{0}=0.8$ (Red), $M_{0}=1M_{\bigodot}$ and
$R=(0.2)^{-1}M_{\bigodot}$ for solution-II.}
\end{figure}
\begin{figure}\center
\epsfig{file=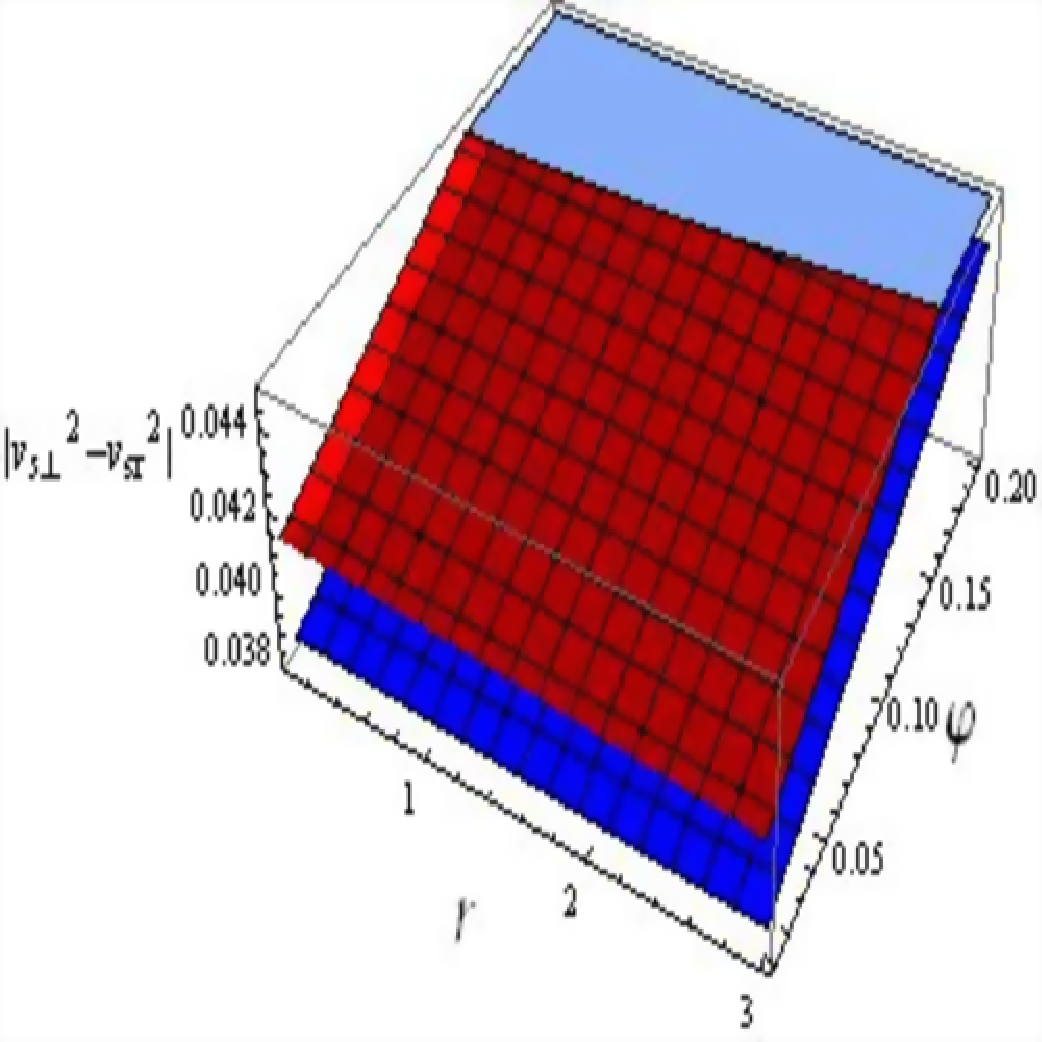,width=0.4\linewidth}\epsfig{file=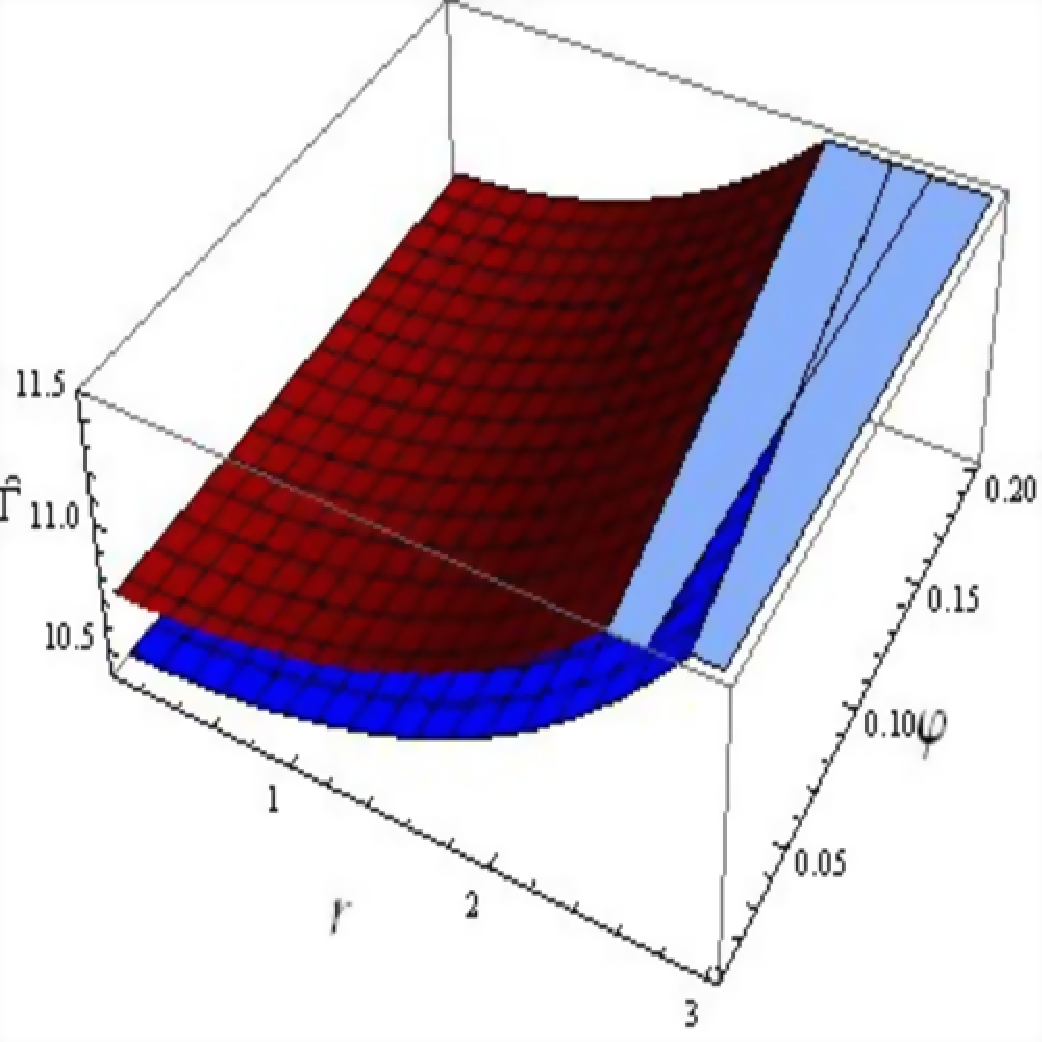,width=0.4\linewidth}
\caption{Plots of $|v_{s\bot}^2-v_{sr}^2|$ and adiabatic index
versus $r$ and $\varphi$ with $S_{0}=0.1$ (Blue), $S_{0}=0.8$ (Red),
$M_{0}=1M_{\bigodot}$ and $R=(0.2)^{-1}M_{\bigodot}$ for
solution-II.}
\end{figure}

\section{Conclusions}

In this paper, we have studied the effects of electromagnetic field
on anisotropic solutions for spherical self-gravitating matter
distribution by means of gravitational decoupling approach in
$f(\mathcal{R},\mathcal{T},\mathcal{Q})$ theory. To overcome the
complex situation in modified gravitational equations, we have
considered a linear model of the form $\mathcal{R}+\beta
\mathcal{Q}$ of this non-minimally coupled theory. We have obtained
two anisotropic solutions by introducing new source
$\Lambda_{\gamma\upsilon}$ in the isotropic fluid. We have assumed
the isotropic Krori-Barua ansatz in modified gravity and calculated
unknown constants through junction conditions. We have employed an
additional constraint on $\Lambda_{\gamma\upsilon}$ to reduce the
number of unknowns in the second sector \eqref{g21}-\eqref{g23}.

We have used constraints on effective pressure and effective energy
density of the fluid distribution leading to solutions-I and II,
respectively. We have examined physical nature of the effective
matter variables $(\hat{\mu},\hat{P}_{r},\hat{P}_{\bot})$,
anisotropic pressure $(\hat{\Delta})$ and energy bounds \eqref{g50}
for $\beta=-0.1$ and $-0.05$ to assess the viability of resulting
solutions in the presence of charge. Both anisotropic solutions meet
the required limit for redshift and compactness factors. For the
first anisotropic solution, the celestial object becomes more dense
with rise in decoupling parameter $\varphi$, whereas the solution-II
provides less dense star. It is also shown that the star becomes
less dense with the increment in charge for both solutions. We have
investigated the stability of obtained solutions through two
different approaches. Both solutions have been found to be
physically viable as well as stable. The stability of solution-I
increases with increase in charge, while solution-II becomes less
stable. It is important to mention here that these solutions in
$f(\mathcal{R},\mathcal{T},\mathcal{Q})$ gravity are physically
feasible and stable for higher values of $\varphi$ in contrast to GR
\cite{35} and $f(\mathcal{G})$ theory \cite{36} in which the second
solution is not stable. For smaller values of $\varphi$, these
results are consistent with $f(\mathcal{R})$ theory \cite{36a}. It
is worthwhile to mention here that all of our results reduce to GR
by taking $f(\mathcal{R},\mathcal{T},\mathcal{Q})=\mathcal{R}$.

\vspace{0.25cm}

\section*{Appendix A}

The matter components in the field equations \eqref{g8}-\eqref{g10}
which contain modified corrections are given as
\begin{eqnarray}\nonumber
\mathcal{T}_{0}^{0(\mathcal{D})}&=&\frac{1}{f_{\mathcal{R}}+\mu
f_{\mathcal{Q}}}\left[\mu\left\{f_{\mathcal{Q}}\left(\frac{\xi'^2}{2e^{\chi}}-\frac{\xi'}{re^{\chi}}+\frac{\xi'\chi'}{4e^{\chi}}
-\frac{\xi''}{2e^{\chi}}-\frac{1}{2}\mathcal{R}\right)+f'_{\mathcal{Q}}\left(\frac{\xi'}{2e^{\chi}}\right.\right.\right.\\\nonumber
&-&\left.\left.\frac{\chi'}{4e^{\chi}}+\frac{1}{re^{\chi}}\right)+\frac{f''_{\mathcal{Q}}}{2e^{\chi}}-2f_{\mathcal{T}}\right\}
+\mu'\left\{f_{\mathcal{Q}}\left(\frac{\xi'}{2e^{\chi}}
+\frac{1}{re^{\chi}}-\frac{\chi'}{4e^{\chi}}\right)+\frac{f'_{\mathcal{Q}}}{e^{\chi}}\right\}\\\nonumber
&+&\frac{f_{\mathcal{Q}}\mu''}{2e^{\chi}}+P\left\{f_{\mathcal{Q}}\left(\frac{3\chi'^2}{4e^{\chi}}-\frac{2}{r^2e^{\chi}}-\frac{\chi''}{2e^{\chi}}\right)
-f'_{\mathcal{Q}}\left(\frac{5\chi'}{4e^{\chi}}-\frac{1}{re^{\chi}}\right)+\frac{f''_{\mathcal{Q}}}{2e^{\chi}}\right\}\\\nonumber
&+&P'\left\{f_{\mathcal{Q}}\left(\frac{1}{re^{\chi}}
-\frac{5\chi'}{4e^{\chi}}\right)+\frac{f'_{\mathcal{Q}}}{e^{\chi}}\right\}+\frac{f_{\mathcal{Q}}P''}{2e^{\chi}}+\frac{\mathcal{R}f_{\mathcal{R}}}{2}
+f'_{\mathcal{R}}\left(\frac{\chi'}{2e^{\chi}}-\frac{2}{re^{\chi}}\right)\\\nonumber
&-&\left.\frac{f''_{\mathcal{R}}}{e^{\chi}}-\frac{f}{2}+\frac{q^2}{r^4}\left\{f_{\mathcal{T}}+\frac{f_{\mathcal{Q}}}{4e^{\chi}}\left(\xi'\chi'
-2\xi''-\xi'^2+\frac{4\chi'}{r}\right)\right\}\right],\\\nonumber
\mathcal{T}_{1}^{1(\mathcal{D})}&=&\frac{1}{f_{\mathcal{R}}+\mu
f_{\mathcal{Q}}}\left[\mu\left(f_{\mathcal{T}}-\frac{f_{\mathcal{Q}}\xi'^2}{4e^{\chi}}
+\frac{f'_{\mathcal{Q}}\xi'}{4e^{\chi}}\right)+\frac{f_{\mathcal{Q}}\mu'\xi'}{4e^{\chi}}+P\left\{f_{\mathcal{T}}
+f_{\mathcal{Q}}\left(\frac{\xi''}{e^{\chi}}\right.\right.\right.\\\nonumber
&-&\left.\left.\frac{\chi'^2}{e^{\chi}}+\frac{\xi'^2}{2e^{\chi}}-\frac{3\xi'\chi'}{4e^{\chi}}-\frac{3\chi'}{re^{\chi}}
+\frac{2}{r^2e^{\chi}}+\frac{1}{2}\mathcal{R}\right)
-f'_{\mathcal{Q}}\left(\frac{\xi'}{4e^{\chi}}+\frac{2}{re^{\chi}}\right)\right\}\\\nonumber
&-&P'f_{\mathcal{Q}}\left(\frac{\xi'}{4e^{\chi}}+\frac{2}{re^{\chi}}\right)+\frac{f}{2}-\frac{\mathcal{R}f_{\mathcal{R}}}{2}
-f'_{\mathcal{R}}\left(\frac{\xi'}{2e^{\chi}}+\frac{2}{re^{\chi}}\right)+\frac{q^2}{r^4}\left\{f_{\mathcal{T}}\right.\\\nonumber
&-&\left.\left.\frac{f_{\mathcal{Q}}}{4e^{\chi}}\left(2\xi''+\xi'^2-\xi'\chi'+\frac{4\xi'}{r}\right)\right\}\right],\\\nonumber
\mathcal{T}_{2}^{2(\mathcal{D})}&=&\frac{1}{f_{\mathcal{R}}+\mu
f_{\mathcal{Q}}}\left[\mu\left(f_{\mathcal{T}}-\frac{f_{\mathcal{Q}}\xi'^2}{4e^{\chi}}
+\frac{f'_{\mathcal{Q}}\xi'}{4e^{\chi}}\right)+\frac{f_{\mathcal{Q}}\mu'\xi'}{4e^{\chi}}+P\left\{f_{\mathcal{T}}
+f_{\mathcal{Q}}\left(\frac{\chi''}{2e^{\chi}}\right.\right.\right.\\\nonumber
&+&\left.\frac{\xi'}{2re^{\chi}}-\frac{3\chi'^2}{4e^{\chi}}-\frac{\chi'}{2re^{\chi}}+\frac{1}{r^2e^{\chi}}-\frac{2}{r^2}
+\frac{1}{2}\mathcal{R}\right)+f'_{\mathcal{Q}}\left(\frac{3\chi'}{2e^{\chi}}-\frac{\xi'}{4e^{\chi}}-\frac{3}{re^{\chi}}\right)\\\nonumber
&-&\left.\frac{f''_{\mathcal{Q}}}{e^{\chi}}\right\}+P'\left\{f_{\mathcal{Q}}\left(\frac{3\chi'}{2e^{\chi}}-\frac{\xi'}{4e^{\chi}}
-\frac{3}{re^{\chi}}\right)-\frac{2f'_{\mathcal{Q}}}{e^{\chi}}\right\}-\frac{f_{\mathcal{Q}}P''}{e^{\chi}}
-\frac{\mathcal{R}f_{\mathcal{R}}}{2}+\frac{f}{2}\\\nonumber
&+&\left.f'_{\mathcal{R}}\left(\frac{\chi'}{2e^{\chi}}-\frac{\xi'}{2e^{\chi}}
-\frac{1}{re^{\chi}}\right)-\frac{f''_{\mathcal{R}}}{e^{\chi}}\right].
\end{eqnarray}
The correction term $\Omega$ in Eq.\eqref{g12} is
\begin{align}
\nonumber \Omega &=
\frac{2}{\left(\mathcal{R}f_{\mathcal{Q}}+2(1+f_{\mathcal{T}})\right)}\left[f'_{\mathcal{Q}}e^{-\chi}\left(P-\frac{q^2}{8\pi
r^4}\right)\left(\frac{1}{r^2}-\frac{e^\chi}{r^2}
+\frac{\xi'}{r}\right)+f_{\mathcal{Q}}e^{-\chi}\right.\\\nonumber
&\times\left(P-\frac{q^2}{8\pi
r^4}\right)\left(\frac{\xi''}{r}-\frac{\xi'}{r^2}-\frac{\chi'}{r^2}-\frac{\xi'\chi'}{r}-\frac{2}{r^3}+\frac{2e^\chi}{r^3}\right)+\left(P'-\frac{qq'}{4\pi
r^4}+\frac{q^2}{2\pi r^5}\right)\\\nonumber
&\times\left\{f_{\mathcal{Q}}e^{-\chi}\left(\frac{\xi'\chi'}{8}
-\frac{\xi''}{8}-\frac{\xi'^2}{8}+\frac{\chi'}{2r}+\frac{\xi'}{2r}+\frac{1}{r^2}-\frac{e^{\chi}}{r^2}\right)+\frac{3}{4}f_{\mathcal{T}}\right\}
+\left(P-\frac{q^2}{8\pi r^4}\right)\\\nonumber &\times
f'_{\mathcal{T}}-\left(\mu+\frac{q^2}{8\pi r^4}\right)
f'_{\mathcal{T}}-\left(\mu'+\frac{qq'}{4\pi r^4}-\frac{q^2}{2\pi
r^5}\right)\left\{\frac{f_{\mathcal{Q}}e^{-\chi}}{8}\left(\xi'^2-\xi'\chi'+2\xi''\right.\right.\\\nonumber
&+\left.\left.\frac{4\xi'}{r}\right)
+\frac{3f_{\mathcal{T}}}{2}\right\}-\left(\frac{e^{-\chi}}{r^2}-\frac{1}{r^2}+\frac{\xi'e^{-\chi}}{r}\right)\left\{\left(\mu'+\frac{qq'}{4\pi
r^4}-\frac{q^2}{2\pi
r^5}\right)f_{\mathcal{Q}}+f'_{\mathcal{Q}}\right.\\\nonumber
&\times\left.\left.\left(\mu+\frac{q^2}{8\pi
r^4}\right)\right\}\right].
\end{align}

\end{document}